\documentstyle[12pt]{article}

\begin{document}
\hfuzz=10pt

\font\twelvemb=cmmib10 scaled \magstep1
\font\tenmb=cmmib10
\font\ninemb=cmmib9
\font\sevenmb=cmmib7
\font\sixmb=cmmib6
\font\fivemb=cmmib5
\font\twelvesyb=cmbsy10 scaled \magstep1
\font\tensyb=cmbsy10
\font\sstwelve=cmss12
\font\ssnine=cmss9
\font\sseight=cmss8
\textfont9=\twelvemb
\scriptfont9=\tenmb
\scriptscriptfont9=\sevenmb
\textfont10=\twelvesyb
\def\bm{\fam9}
\def\bms{\fam10}

\newcommand{\DS}{\displaystyle}
\newcommand{\TS}{\textstyle}
\newcommand{\SS}{\scriptstyle}
\newcommand{\SSS}{\scriptscriptstyle}

\newcommand\zZtwelve{\hbox{\sstwelve Z\hskip -4.5pt Z}}
\newcommand\zZnine{\hbox{\ssnine Z\hskip -3.9pt Z}}
\newcommand\zZeight{\hbox{\sseight Z\hskip -3.7pt Z}}
\newcommand\zZ{\mathchoice{\zZten}{\zZten}{\zZeight}{\zZeight}}
\newcommand\ZZ{\mathchoice{\zZtwelve}{\zZtwelve}{\zZnine}{\zZeight}}

\mathchardef\sigma="711B
\mathchardef\tau="711C
\mathchardef\omega="7121
\mathchardef\nabla="7272

\newcommand{\e}{\epsilon}
\newcommand{\eps}{\epsilon}
\newcommand{\ee}{\varepsilon}
\newcommand{\vp}{\varphi}
\newcommand{\vphi}{\varphi}
\newcommand{\cphi}{\Phi}
\let\oldvrho=\varrho
\newcommand{\vrho}{{\raise 2pt\hbox{$\oldvrho$}}}
\let\oldchi=\chi
\renewcommand{\chi}{{\raise 2pt\hbox{$\oldchi$}}}
\let\oldxi=\xi
\renewcommand{\xi}{{\raise 2pt\hbox{$\oldxi$}}}
\let\oldzeta=\zeta
\renewcommand{\zeta}{{\raise 2pt\hbox{$\oldzeta$}}}

\newcommand{\la}[1]{\label{#1}}
\newcommand{\ur}[1]{(\ref{#1})}
\newcommand{\ra}[1]{(\ref{#1})}
\newcommand{\urs}[2]{(\ref{#1},~\ref{#2})}
\newcommand{\eq}[1]{eq.~(\ref{#1})}
\newcommand{\eqs}[2]{eqs.~(\ref{#1},~\ref{#2})}
\newcommand{\eqss}[3]{eqs.~(\ref{#1},~\ref{#2},~\ref{#3})}
\newcommand{\eqsss}[2]{eqs.~(\ref{#1}--\ref{#2})}
\newcommand{\Eq}[1]{Eq.~(\ref{#1})}
\newcommand{\Eqs}[2]{Eqs.~(\ref{#1},~\ref{#2})}
\newcommand{\Eqss}[3]{Eqs.~(\ref{#1},~\ref{#2},~\ref{#3})}
\newcommand{\Eqsss}[2]{Eqs.~(\ref{#1}--\ref{#2})}
\newcommand{\fig}[1]{Fig.~\ref{#1}}
\newcommand{\figs}[2]{Figs.~\ref{#1},\ref{#2}}
\newcommand{\figss}[3]{Figs.~\ref{#1},\ref{#2},\ref{#3}}
\newcommand{\beq}{\begin{equation}}
\newcommand{\eeq}{\end{equation}}

\newcommand{\doublet}[3]{\:\left(\begin{array}{c} #1 \\#2
            \end{array} \right)_{#3}}
\newcommand{\vect}[2]{\:\left(\begin{array}{c} #1 \\#2
            \end{array} \right)}
\newcommand{\vectt}[3]{\:\left(\begin{array}{c} #1 \\#2 \\ #3
            \end{array} \right)}
\newcommand{\vectf}[4]{\:\left(\begin{array}{c} #1 \\#2 \\#3\\#4
            \end{array} \right)}
\newcommand{\matr}[4]{\left(\begin{array}{cc}
                   #1 &#2 \\
                   #3 &#4 \end{array} \right)}
\newcommand{\fracsm}[2]{{\textstyle\frac{#1}{#2}}}

\newcommand{\D}{{\cal D}}
\newcommand{\K}{{\cal K}}
\newcommand{\NC}{N_{\rm CS}}
\newcommand{\Ncs}{N_{\rm CS}}
\newcommand{\SU}{$SU(2)~$}
\newcommand{\Pmax}{P_{max}}
\newcommand{\tr}{\,{\rm tr}\,}
\newcommand{\Tr}{\,{\rm Tr}\,}
\newcommand{\ldef}{=}
\newcommand{\rdef}{=}
\newcommand{\simlt}{\stackrel{<}{{}_\sim}}
\newcommand{\simgt}{\stackrel{>}{{}_\sim}}

\newcommand{\nuH}{\nu_{\SSS H}}
\newcommand{\nuF}{\nu_{\SSS F}}
\newcommand{\nuf}{\nu_{\SSS f}}
\newcommand{\nut}{\nu_{\SSS t}}
\newcommand{\nuHold}{\nu_{{\SSS H}\,{\rm old}}}

\newcommand{\op}[1]{{\bf \hat{#1}}}
\newcommand{\opr}[1]{{\rm \hat{#1}}}
\newcommand{\bra}[1]{\langle#1\vert}
\newcommand{\ket}[1]{\vert#1\rangle}
\newcommand{\lsim}{\mathrel{\lower 2pt\hbox{$\stackrel{<}{\SS\sim}$}}}
\newcommand{\gsim}{\mathrel{\lower 2pt\hbox{$\stackrel{>}{\SS\sim}$}}}

\def\theequation{\arabic{section}.\arabic{equation}}
\def\appendix{\par
 \setcounter{section}{0}
 \def\thesection{Appendix}
 \def\theequation{\Alph{section}.\arabic{equation}}}

{\thispagestyle{empty}
\rightline{RUB-TPII-25/95}
\vspace{2cm}
\begin{center}
{\Large\bf
Sphaleron transitions in the Minimal Standard Model
and the upper bound for the Higgs Mass\\}
\vspace{27 pt}
{\large\bf
Dmitri Diakonov$^*$\footnote{
\noindent
diakonov@lnpi.spb.su},
Maxim Polyakov$^*$,
Peter Sieber$^\diamond$, \\ J\"org Schaldach$^\diamond$,
and Klaus Goeke$^\diamond$\footnote{
\noindent
goeke@hadron.tp2.ruhr-uni-bochum.de}}

\vspace{20 pt}
{\small\it $^*$Petersburg
Nuclear Physics Institute, Gatchina, St.~Petersburg 188350, Russia \\
$^\diamond$Institut f\"ur Theor.~Physik I\hskip-1.5pt I, Ruhr-Universit\"at
Bochum, D-44780 Bochum, Germany}
\end{center}
\vspace{30 pt}
\abstract{We calculate the dissipation of the baryon number after the
electroweak phase transition due to thermal fluctuations above the
sphaleron barrier. We consider not only the classical Boltzmann
factor but also fermionic and bosonic one-loop contributions.
We find that both bosonic and especially fermionic
fluctuations can considerably suppress the transition rate. Assuming the
Langer--Affleck formalism for this rate, the
condition that an initial baryon asymmetry must not be washed
out by sphaleron transitions
leads, in the Minimal Standard Model ($\sin\theta_W=0$),
to an upper bound for the Higgs
mass in the range 60 to 75 GeV.}}

\newpage

\section{Introduction}
\setcounter{equation}{0}

The question about the origin of the baryon asymmetry of
the Universe (BAU) has recently gained much interest.
Many different models of how the
BAU was created are being discussed in the literature
(for reviews see e.g.~\cite{Dolgov}),
some of them considering BAU generation at the GUT stage of the
Universe, others favoring the generation during the electroweak
phase transition.

Whatever the mechanism was which led to the BAU at early times,
the resulting asymmetry might
have been eliminated by baryon number violating processes in the electroweak
theory after the phase transition. Such processes are
possible due to the anomaly  of baryon and lepton currents \cite{tHooft}
and the non-trivial topological structure of the
Yang--Mills theory. This feature was discovered in 1976 by Faddeev
\cite{Faddeev} and Jackiw and Rebbi \cite{Jackiw}, who found that the
potential energy is periodic in a certain functional of the fields, the
Chern--Simons number $\NC$.
Topologically distinct
vacua of the theory are enumerated by integer $\NC$.
In the electroweak theory those
vacua are separated by an energy barrier whose height is of the order of
$m_W/\alpha$
where $m_W$ is the mass of the $W$ boson
and $\alpha=g^2/(4\pi)$ is the $SU(2)$ gauge coupling
constant.

Transitions from one vacuum to a topologically distinct one over this
barrier change the baryon and lepton number by one unit per
fermion generation due to the  anomaly of
the corresponding currents.  If we assume in accordance with the
standard model that $B-L$ (baryon minus lepton number)
is conserved and that there is no primordial
excess of say, antileptons, then these transitions can cause the BAU
erasure as mentioned above.  Hence it is necessary to know the
transition rate of such processes.  While the baryon number of the
Universe today ($B_0$) is about $10^{-9}$ to $10^{-10}$ (relative to
the number of relict photons), in models generating the BAU at the
electroweak phase transition the number of produced baryons ($B_{T_c}$)
per photon is of order $10^{-5}$ \cite{Boch,Shap}.
(The precise value is not
important for our calculation, see below). Thus the ratio $B_0/B_{T_c}$
which describes the dissipation of the BAU should not be significantly
lower than about $10^{-5}$, otherwise the initial baryon excess is not
large enough to explain the present day BAU.

In principle to obtain the value of this ratio one has to integrate
the rate of the baryon number violating processes over the temperature
from $T=T_c$ to $T=0$. In practice, however, the
rate is very strongly suppressed at ordinary temperatures \cite{tHooft}
so that only a short range below $T_c$ contributes to
the erasure of the BAU. While at low $T$ the rate is dominated
by tunnelling processes, at higher temperatures the energy barrier
can be overcome thanks to thermal fluctuations \cite{Kuzmin,Arnold,Boch}.
This thermal transition rate can
be evaluated by the semi-classical formalism of Langer \cite{Langer} and
Affleck \cite{Affleck}.

A key role in this calculation is played by the static classical field
configuration which corresponds to the top of the energy barrier,
having Chern--Simons number $\NC=\fracsm{1}{2}$.
This configuration was first found by Dashen,
Hasslacher and Neveu \cite{Dashen} and rediscovered in the context of
electroweak theory by Klinkhamer and Manton, it is called sphaleron
\cite{Klinkhamer}. Its energy $E_{\rm class}$ enters the transition rate
$\gamma(T)=A(T)\,e^{-E_{\rm class}/T}$ via the classical Boltzmann
factor and is usually the dominant contribution
to $\gamma$, i.e.~in most cases $|\ln A(T)| < E_{\rm class}/T$.
The prefactor $A(T)$ contains contributions coming from fermion and
boson quantum fluctuations about the sphaleron.
In \cite{Arnold} the rate was calculated
considering the classical and zero-mode contributions,
while the determinant of non-zero boson fluctuation modes
and the fermion determinant were set to unity.  The result for $\gamma$
was so large that any initial baryon excess would have been
washed out by sphaleron transitions after the phase transition.
Therefore it is of interest whether quantum loop corrections
could help to preserve the baryon asymmetry.

Several investigations which consider loop corrections have already been
made. Bochkarev and Shaposhnikov \cite{Boch,Shap} included
the boson fluctuations through an effective potential of the
Higgs field. They obtained that the transition rate is sufficiently
suppressed if
the Higgs mass is below an upper limit of 45 to 55 GeV.
A direct computation of the bosonic determinant over non-zero
modes was made in \cite{McLCar} by using an approximation technique
\cite{DPY}, exact calculations were performed in \cite{McLetal,Baacke}.

All these calculations were based on the
high temperature limit in which the four-dimensional fluctuation matrix
can be replaced by the three-dimensional one and fermions decouple
completely. Although parametrically this limit is reasonable
it need not necessarily be justified numerically.
In this paper we go beyond the high temperature approximation
which corresponds to taking into account the fermion determinant
(suppressed in the formal high temperature limit) and
to calculating the full four-dimensional bosonic determinant.
We generalize the preceeding
calculations to arbitrary temperatures. First, we include
fermion loops which previously have been altogether neglected, second,
we evaluate the fluctuation determinants
for finite temperature considering the full sum over Matsubara frequencies.
Both contributions seem to be quite essential numerically.

The fermion determinant has been considered in detail in our previous
publication \cite{rub}, the bosonic one is the aim of this work
(see also \cite{rub2}).
We find that the result is significantly influenced
by terms which vanish in the formal high $T$ limit, especially by the
contribution of the fermion fluctuations.
Actually, the polarization of the Dirac sea of fermions in the classical
sphaleron background field adds up to about $30\,\%$ to the sphaleron energy.
Therefore, the fermion determinant which was put to unity in
\cite{Boch,Arnold,McLCar,McLetal,Baacke} leads to a strong
additional suppression of the transition rate.

Let us remark that in an abelian (1+1)-dimensional model the
transition rate can be calculated analytically. This has been done in
\cite{Boch1} for the boson and in \cite{Gould} for the fermion loop
correction. In (3+1) dimensions, however,
one has to resort to numerical methods.

For the evaluation of the boson determinant we use the same method as
for the fermion determinant in \cite{rub}.
It is based on the computation of the complete (discretized) spectrum of
the fluctuation operators.
All the relevant quantities can subsequently be calculated for any
temperature $T$ by suitable
summations over the eigen-energies. Including the loop corrections into
the formula for the transition rate, we finally obtain the ratio
$B_0/B_{T_c}$ which is a measure for the erasure of the BAU. We find
that both bosonic and fermionic fluctuations suppress the rate
considerably, especially for a low mass of the Higgs boson and a large
top quark mass. For a top quark mass in the range $m_t=150$ to 200 GeV,
in accordance to recent experimental results \cite{topmass},
the condition that the BAU must not be washed out by
sphaleron transitions leads within the framework of our one-loop calculation
and the Langer--Affleck formalism to an upper limit for $m_H$ in the
range between about 60 and 75 GeV.

Another goal of this work is the
recalculation of the boson fluctuation determinant in the high $T$
limit since
the results of the two existing calculations \cite{McLetal,Baacke},
based on different analytical and numerical techniques,
deviate from each other. Although they show the same
tendency for low Higgs masses, no satisfactory quantitative agreement
was found. Our numerical method is based on the diagonalization of the
fluctuation operator as described above; it differs significantly
from those used in \cite{McLetal,Baacke} so that our study can be
considered as independent. We find that
our results for the boson determinant in the high $T$ limit agree
with the results of \cite{Baacke} up to
about $10\%$ while there is a larger deviation from the ones of
\cite{McLetal}.

The paper is organized as follows: In Section 2 we introduce into
the notations and conventions of the model,
in Section 3 we apply the Langer--Affleck
formalism to the baryon number violation processes.
The renormalization is discussed in Section 4,
followed by the treatment of the temperature dependent parts of the
fluctuations.
The evaluation of the baryon erasure $B_0/B_{T_c}$ is
done in Section 6. Numerical results and checks for the
computation of the fluctuation determinants are presented in
Section 7. In Section 8 we proceed to the evaluation of the
sphaleron transition rate and deduce the upper bound for the
Higgs mass. We also investigate the applicability of the
framework of our calculation. Finally we summarize the results
and draw our conclusions in Section 9.
Technical details about the computation of the discretized
spectrum and the spectral densities are treated in the appendices.

\section{The model and parameters}
\setcounter{equation}{0}

We consider the minimal version of the standard electroweak theory
with one Higgs doublet which is Yukawa coupled to left handed
fermion doublets and to right handed singlets; in the following we
write only one doublet and one pair of singlets for brevity.

We shall work in the limit of the vanishing Weinberg angle,
i.e.~the theory is reduced to the pure $SU(2)$ case
(without  the $U(1)$).
This idealization does not seem to be significant \cite{Kunz}.
The Lagrangian is thus
\begin{eqnarray}
{\cal L} &=&
{}-\frac{1}{4g^2}F_{\mu\nu}^a F^{a\,\mu\nu}
+ (D_\mu \Phi)^\dagger(D^\mu \Phi) - \frac{\lambda^2}{2}
\Bigl(\Phi^\dagger \Phi-\frac{v^2}{2}\Bigr)^2 \nonumber \\
 & &{}+\bar{\psi}_L i\gamma^\mu D_\mu \psi_L +
      \bar{\psi}_R i\gamma^\mu \partial_\mu \psi_R
- \bar{\psi}_L M \psi_R - \bar{\psi}_R M^\dagger\psi_L\quad\quad
\la{Lagra}\end{eqnarray}
with the covariant derivative
$D_\mu=\partial_\mu -i A_\mu\;,\ A_\mu=\frac{1}{2}A_\mu^a\tau^a$,
and the field strength $F_{\mu\nu}=\frac{1}{2}F_{\mu\nu}^a\tau^a=
i[D_\mu,D_\nu]\;,\ F_{\mu\nu}^a=\partial_\mu A_\nu^a-\partial_\nu A_\mu^a
+\e^{abc}A_\mu^b A_\nu^c$.
$M$ is a $2\times 2$ matrix built of the Higgs field components
$\Phi={\cphi^+ \choose \cphi^0}$
and the Yukawa couplings $h_u, h_d$:
\beq
M=\matr{h_u \cphi^{0\ast}}{h_d \cphi^+}{- h_u \cphi^{+\ast}}{h_d \cphi^0}.
\la{MM}\eeq
$\psi_L$ means the \SU fermion doublet
\beq
\psi_L = \fracsm{1}{2}(1-\gamma_5)\psi = {\psi_L^u\choose\psi_L^d}
\eeq
and with $\psi_R$ we denote the pair of the singlets
\beq
\psi_R = \fracsm{1}{2}(1+\gamma_5)\psi = {\psi_R^u\choose\psi_R^d}\;.
\eeq
The masses generated by the non-vanishing vacuum expectation value
$\bra{0}\Phi\ket{0}=\fracsm{v}{\sqrt{2}}{0\choose 1}$ are
\beq
m_W=\frac{gv}{2}\;,\qquad
m_{u,d}=\frac{h_{u,d}v}{\sqrt{2}}\;,\qquad\mbox{and}\qquad
m_H=\lambda v
\la{masses}\eeq
for the gauge boson, the fermions, and the Higgs boson.

We prefer to work in terms of dimesionless rescaled quantities
\beq
x^\mu\to m_W^{-1}x^\mu\;,\qquad
A^a_\mu\to m_W A^a_\mu\;,\qquad
\Phi\to\frac{m_W}{\sqrt{2}\,g}\Phi\;;
\nonumber\eeq
using the following representation of the Dirac matrices
\beq
\gamma^0=\matr{0}{1}{1}{0}
\;,\qquad
\gamma^i=\matr{0}{\sigma_i}{-\sigma_i}{0}\;,\qquad
\gamma^5=\matr{-1}{0}{0}{1}\;,
\eeq
the fermion Dirac spinors can be reduced to two components:
\beq
\psi_{L}\to m_W^{3/2}{\psi_{L}\choose 0}\;,\qquad
\psi_{R}\to m_W^{3/2}{0\choose \psi_{R}}\;.
\la{scalea}
\eeq
In this representation the Lagrangian \ur{Lagra} is
\begin{eqnarray}
{\cal L}&=&m_W^4\Biggl[\frac{1}{g^2}\biggl(
-\fracsm{1}{4}F_{\mu\nu}^a F^{a\,\mu\nu}
+\fracsm{1}{2}(D_\mu \Phi)^\dagger(D^\mu \Phi)
- \fracsm{1}{32}\nuH^2(\Phi^\dagger\Phi-4)^2\biggr)\la{Lagrasc}\\
 & &\quad{}+i\psi_L^\dagger(D_0-\sigma_iD_i)\psi_L
+i\psi_R^\dagger(\partial_0+\sigma_i\partial_i)\psi_R
- \psi_L^\dagger M \psi_R - \psi_R^\dagger M^\dagger\psi_L\Biggr]\;,\qquad
\nonumber\end{eqnarray}
with the mass matrix
\beq
M=\frac{1}{2m_W}
\matr{m_u\cphi^{0\ast}}{m_d\cphi^+}{-m_u\cphi^{+\ast}}{m_d\cphi^0}
\qquad\mbox{and}\qquad
 \nuH=\frac{m_H}{m_W}\;.
\la{Mkap}\eeq

\section{The baryon number violation rate}
\setcounter{equation}{0}

As it is well known \cite{tHooft} the baryon and lepton
numbers $B$ and $L$
are not conserved in the standard model of electroweak
interactions due to the anomaly of the corresponding currents
$j_B^\mu$ and $j_L^\mu$:
\beq
\partial_\mu j_B^\mu = \partial_\mu j_L^\mu
= \frac{N_g}
{64\pi^2} \epsilon^{\mu\nu\rho\sigma} F_{\mu\nu}^a  F_{\rho\sigma}^a.
\eeq
Here $N_g=3$ is the number of fermion generations.
We dropped the contribution of the $U(1)$ gauge field
to the anomaly since we work in the approximation
of vanishing Weinberg angle $\sin \theta_W=0$.

Integrating this anomaly equation one finds
that in any process the change of the baryon and lepton
numbers is related to the change of the Chern--Simons number by
\beq
\Delta B = \Delta L = N_g\,\Delta \Ncs\;,
\eeq
where the Chern--Simons number $\Ncs$ is defined by
\beq
\Ncs = \int d^3 {\bf r}\,K_0({\bf r})\,, \la{NCS}
\eeq
\beq
K^\mu = \frac1{16\pi^2}\,\epsilon^{\mu\nu\rho\sigma}
\left(A_\nu^a \partial_\rho A_\sigma^a
+ \fracsm{1}{3} \epsilon_{abc} A_\nu^a A_\rho^b A_\sigma^c
\right),
\eeq
\beq
\partial_\mu K^\mu = \frac1{64\pi^2}
\epsilon^{\mu\nu\rho\sigma} F_{\mu\nu}^a  F_{\rho\sigma}^a\;.
\eeq
The semiclassical description of the processes with fermion and
lepton number violation is based on the existence of an infinite
number of classical vacuum configurations labeled by integer
values of $\Ncs$. These vacua are separated by potential barriers
which can be overcome either by quantum tunnelling or
by real processes in the Minkowskian time due to thermal fluctuations.
We shall be interested in temperatures at which the
real-time transitions dominate. The rate of the thermal transitions
at temperature $T$ between adjacent vacua is roughly given
by the Boltzmann factor $\exp(-E_{\rm class}/T)$, where $E_{\rm class}$
is the energy of the sphaleron, which is the field configuration at
the top of the barrier between the two vacua.
The transition rate $\Gamma$ with the preexponential
factor is given by the Langer--Affleck formula \cite{Langer,Affleck}
\beq
\Gamma = \frac{|\omega_-|}\pi \, \frac{{\rm Im}\,Z_{\rm sphal}}{Z_0}\;.
\la{Gamma-rate-1}
\eeq
Here $Z_0$ is the partition function computed in the semiclassical
approximation around the vacuum, and $Z_{\rm sphal}$ is the partition
function obtained by semiclassical expansion about the sphaleron
solution. Since the sphaleron solution is a saddle-like point
of the potential energy functional, the quadratic form of fluctuations
about the sphaleron has a negative mode $\omega_-^2 < 0$
so that the formal semiclassical expansion about the non-stable static
solution  gives a complex contribution to the partition function
$Z_{\rm sphal}$.

In the Weinberg--Salam model the sphaleron solution
in the temporal gauge $A_0=0$
can be found as a stationary point of the energy functional
\beq
E_{\rm class}=\frac{m_W}{g^2}\int
d^3{\bf r}\,\left[\fracsm{1}{4}(F^a_{ij})^2+\fracsm{1}{2}(D_i\Phi)^\dagger
(D_i\Phi)+\fracsm{1}{32}\nuH^2(\Phi^\dagger\Phi-4)^2\right]\;.
\la{Eclass}
\eeq
which leads to the classical equations
\begin{eqnarray}
& &(  D_i  F_{ij})^a-\fracsm{i}{4}\Bigl( \Phi^\dagger
\tau^a(  D_j \Phi)-(  D_j \Phi)^\dagger\tau^a \Phi\Bigr)=0
\nonumber\\
& &\Bigl(  D_i^2-\fracsm{1}{8}\nuH^2
\bigl( \Phi^\dagger \Phi-4\bigr)\Bigr) \Phi=0\;.
\la{motion}\end{eqnarray}
The sphaleron solution is assumed to have the following form (hedgehog)
\begin{eqnarray}
\bar A^a_i({\bf r})&=&\epsilon_{aij}
n_j\,\frac{1-A(r)}{r}+(\delta_{ai}-n_an_i)\,
\frac{B(r)}{r}+n_an_i\,\frac{C(r)}{r}\;,\nonumber\\
\bar\Phi({\bf r})&=&2\Bigl[H(r)+i G(r)\,{\bf n}\cdot{\bm\tau}\Bigr]
{\textstyle{0 \choose 1}}\;.
\la{HeHo}
\end{eqnarray}
Here ${\bf r}=r{\bf n}$, ${\bm\tau}$ are Pauli matrices,
and the profile functions $A$,$B$,$C$,$G$,$H$ can be found
numerically by solving the classical equations.

The spherical symmetry of this static solution is preserved under
time-in\-de\-pen\-dent gauge transformations of the form
\beq
U({\bf r})=\exp\,\Bigl[iP(r)\,{\bf n}\cdot{\bm\tau}\Bigr]
\la{HeHotr}
\eeq
with an arbitrary function $P(r)$.
One of the five profile functions could be completely eliminated by
using this gauge freedom, but since in this case the remaining functions
are not necessarily regular at the origin and at infinity, which is
required by our numerics, we use all five functions.

It should be mentioned
that the expression for the Chern--Simons Number given in \eq{NCS}
is not gauge invariant, so it is only well-defined if we require
the fields to be continuous at infinity. In this case $\Ncs$ is determined
up to an integer number which is the winding number of a possible gauge
transformation. The final results
should be independent of the choice of the gauge. Since we did not exploit
the gauge freedom to eliminate one of the five profile functions, we can
verify this gauge invariance numerically. This provides a powerful
non-trivial check of our performance.

We can rewrite the general Langer--Affleck formula \eq{Gamma-rate-1}
in the form
\beq
\Gamma=\frac{m_W\omega_-}{2\pi}\;
\frac{\DS \int\D a\,\D\vphi\,\D\psi^\dagger\D\psi\,\exp\Bigl(-S\bigl[
\bar A+ga,\bar\Phi+g\vphi,\psi^\dagger,\psi\bigr]
{\TS{\rm 2^{nd}\ \,\atop order}}\;\Bigr)}
{\DS \int\D a\,\D\vphi\,\D\psi^\dagger\D\psi\,\exp\Bigl(-S\bigl[
\bar A^{(0)}+ga,\bar\Phi^{(0)}+g\vphi,\psi^\dagger,\psi\bigr]
{\TS{\rm 2^{nd}\ \,\atop order}}\;\Bigr)}\ .
\la{LA1}\eeq
Here the Euclidean action
\begin{eqnarray}
\lefteqn{
S[A,\Phi,\psi^\dagger,\psi]=\frac{1}{g^2}\int_0^{\beta m_W}dt
\int d^3{\bf r}\biggl[\fracsm{1}{4} (F_{\mu\nu}^a)^2
+\fracsm{1}{2}(D_\mu\Phi)^\dagger(D_\mu\Phi)}\la{Seuc}\\
&&{}+\fracsm{1}{32}\nuH^2(\Phi^\dagger\Phi-4)^2
+g^2\,\bigl(\psi_L^\dagger\ \psi_R^\dagger\bigr)
\matr{D_0+i\sigma_iD_i}{M}{M^\dagger}
{\partial_0-i\sigma_i\partial_i}{\psi_L\choose\psi_R}\biggr]\nonumber
\end{eqnarray}
is expanded to second order in fluctuations around the sphaleron
configuration $\bar A,\bar\Phi$ in the numerator and around the vacuum
configuration $\bar A^{(0)},\bar\Phi^{(0)}$ in the denominator.
$\beta=1/T$ is the inverse of the temperature $T$.

In zeroth order (no fluctuations) $S$ reduces to $\beta$ times the classical
energy \ur{Eclass} of the sphaleron or the vacuum configuration, hence the
transition rate $\Gamma$ contains the Boltzmann factor
$\,\exp\bigl[-\beta(E_{\rm class}-E_{\rm class}^{(0)})\bigr]$ mentioned above,
where we have $E_{\rm class}^{(0)}= E_{\rm class}[\bar A^{(0)},
\bar\Phi^{(0)}]=0$.

In fact, formula \ur{LA1} should be modified to take into account
the gauge fixing, renormalization and special treatment of zero
modes. Let us start with the gauge fixing.
Following \cite{McLetal} we shall work in the
background $R_\oldxi$ gauge defined by adding the term
\beq
  \fracsm{1}{2}\,\Bigl((\bar D_\mu a_\mu)^a
   +\fracsm{i}{4}\bigl(\bar\Phi^\dagger\tau^a\vphi
     -\vphi^\dagger\tau^a\bar\Phi\bigr)\Bigr)^2 \la{gfix}
\eeq
to the quadratic form for the fluctuations $a$, $\vphi$.
$\bar D$ is the covariant derivative with the background
field $\bar A$.
In this gauge the Faddeev--Popov determinant is
\beq
\kappa_{\rm FP}^2=
|\det\Bigl(-\partial_0^2+\K_{\rm FP}\Bigr)|
\qquad\mbox{with}\qquad
\K_{\rm FP}=
-\bar D_i^2+\fracsm{1}{4}\,\bar\Phi^\dagger\bar\Phi\;.
\la{kFP}\eeq
The quadratic form of the action expanded about the sphaleron solution
in this gauge is
\begin{eqnarray}
\delta^{(2)} S&=&\fracsm{1}{2}\int_0^{\beta m_W}d^4x\,\biggl[
a^a_\mu\Bigl(-\partial_0^2-\bar D_i^2
+\fracsm{1}{4}\bar\Phi^\dagger\bar\Phi\Bigr)^{ab}a^b_\mu
+2\e^{abc}\bar F^c_{ij}a_i^a a_j^b\nonumber\\
& &{}-i\,a_i^a\Bigl((\bar D_i\bar\Phi)^\dagger\tau^a\vphi
-\vphi^\dagger\tau^a(\bar D_i\bar\Phi)\Bigr)
+\frac{\nuH^2}{16}
\bigl(\bar\Phi^\dagger\vphi+\vphi^\dagger\bar\Phi\bigr)^2\nonumber\\
& &{}-\frac{1}{16}\bigl(\bar\Phi^\dagger\tau^a\vphi
-\vphi^\dagger\tau^a\bar\Phi\bigr)^2
+\vphi^\dagger\Bigl(-\partial_0^2-\bar D_i^2+\frac{\nuH^2}{8}
\bigl(\bar\Phi^\dagger\bar\Phi-4\bigr)\Bigr)\,\vphi\la{Seff2}\\
& &{}+2\,\bigl(\psi_L^\dagger\ \psi_R^\dagger\bigr)
\matr{\partial_0+i\sigma_i\bar D_i}{\bar M}{\bar M^\dagger}
{\partial_0-i\sigma_i\partial_i}{\psi_L\choose\psi_R}\biggr],
\nonumber
\label{quadratic-form-1}
\end{eqnarray}
where $\bar M$ is the mass matrix \ur{Mkap} with the background
Higgs field $\bar\Phi$.
The integration over $a_0$ yields the inverse square-root of the
Faddeev--Popov determinant, $1/\kappa_{\rm FP}$.
The spherical symmetry of the sphaleron solution
leads to certain symmetries of the quadratic form of the
action. In order to make these symmetries explicit in the Higgs sector,
we prefer to work with the complex Higgs doublets in terms of four real
components.
Any complex doublet $\xi$ can be represented in the form
\beq
\xi={\xi_2+i\xi_1\choose\xi_4-i\xi_3}=\xi_\mu\tau^+_\mu{0\choose 1}
\qquad\mbox{with}\qquad
\tau^{\pm}_\mu=(\pm i{\bm\tau},1)\qquad(\mu=1,\dots,4)\;;
\eeq
where $\xi_\mu$ is a real four vector.
One has
\begin{eqnarray}
& &\xi^\dagger\zeta+\zeta^\dagger\xi=2\xi_\mu\zeta_\mu\nonumber\\
& &i\,\bigl(\xi^\dagger\tau^a\zeta-\zeta^\dagger\tau^a\xi\bigr)=
2\eta^a_{\mu\nu}\xi_\mu\zeta_\nu
\end{eqnarray}
where $\eta^a_{\mu\nu}$ is  the 't Hooft symbol
\beq
\eta^a_{\mu\nu}=\frac{1}{4i}\tr\Bigl[\tau^a(\tau^+_\mu\tau^-_\nu
-\tau^+_\nu\tau^-_\mu)\Bigr]\;,\qquad
\eta^a_{\mu 4}=-\eta^a_{4\mu}=\delta_{a\mu}\;,\qquad
\eta^a_{ij}=\e_{aij}\;.
\eeq
In this four component language the covariant derivative
$D_\lambda\xi=(D_\lambda\xi)_\mu\tau^+_\mu{0\choose 1}$ can be written as
\beq
(D_\lambda\xi)_\mu=(D_\lambda)_{\mu\nu}\xi_\nu\qquad\mbox{with}\qquad
(D_\lambda)_{\mu\nu}=\delta_{\mu\nu}\partial_\lambda-\fracsm{1}{2}
\eta^a_{\mu\nu}A^a_\lambda\;.
\eeq

Joining the fluctuations of the gauge and Higgs fields
into a $13$ component vector $(a_i^a \> \vphi_\mu )$
we can rewrite the quadratic form of the action
\ur{quadratic-form-1} in the form
\beq
\delta^{(2)} S=\int_0^{\beta m_W}d^4x\,
\biggl[
    \fracsm{1}{2}  ( a \> \varphi) (-\partial_0^2 + \K_{\rm bos} )
    {a \choose \varphi}
    +\,\bigl(\psi_L^\dagger\ \psi_R^\dagger\bigr)
  (\partial_0+{\cal H}_{\rm ferm})   {\psi_L\choose\psi_R}
\biggr],
\label{quadratic-form-2}
\eeq
where
\begin{eqnarray}
&&\K_{\rm bos}=\matr{{\cal G}^{ab}_{ij}}{{\cal W}^a_{i\nu}}
{{\cal W}^b_{j\mu}}{{\cal H}_{\mu\nu}}\;,
\nonumber\\
&&\qquad\quad{\cal G}^{ab}_{ij}=
\delta_{ij}\Bigl(-(\bar D_k^2)^{ab}+\fracsm{1}{4}\,\delta^{ab}
\bar\Phi_\alpha\bar\Phi_\alpha\Bigr)+2\e^{abc}\bar F_{ij}^c\;,\nonumber\\
&&\qquad\quad{\cal H}_{\mu\nu}=
-(\bar D_k^2)_{\mu\nu}-\fracsm{1}{4}
(1-\nuH^2)\bar\Phi_\mu\bar\Phi_\nu+\fracsm{1}{4}\,\delta_{\mu\nu}\Bigl(
(1+\fracsm{1}{2}\nuH^2)\bar\Phi_\alpha\bar\Phi_\alpha-2\nuH^2\Bigr)\;,
\qquad\nonumber\\
&&\qquad\quad{\cal W}^a_{i\nu}=\eta^a_{\nu\alpha}(\bar D_i\bar\Phi)_\alpha
\;,\nonumber\\
&&{\cal H}_{\rm ferm}=\matr{i\sigma_i\bar D_i}
{\fracsm{1}{2}\nuF\bar\Phi_\mu\tau_\mu^+}
{\fracsm{1}{2}\nuF\bar\Phi_\mu\tau_\mu^-}{-i\sigma_i\partial_i}\;.
\la{Kfluc}\end{eqnarray}
In order to preserve the spherical symmetry of the fermionic Hamiltonian
${\cal H}_{\rm ferm}$
we have taken equal masses for up and down fermions:
\beq
m_u=m_d\,, \qquad \nu_F=\frac{m_u}{m_W}=\frac{m_d}{m_W}\;.
\eeq
The physical significance of this approximation will be discussed later.

Using the quadratic form of the action \ur{quadratic-form-2}
we can perform the functional integral in the
Langer--Affleck formula for the transition rate \ra{LA1}
and get \cite{Boch,Arnold}
\beq
\Gamma
=\frac{m_W\omega_-}{2\pi}\,
\frac{8\pi^2V}{(g^2\beta)^3}\,N_{\rm tr}^3N_{\rm rot}^3\,
\frac{\DS\kappa_{\rm FP}}{\DS\kappa_{\rm FP}^{(0)}}\,
\frac{\DS\kappa_{\rm bos}^{(0)}}{\DS\kappa_{\rm bos}}\,
\frac{\DS\kappa_{\rm ferm}}{\DS\kappa_{\rm ferm}^{(0)}}\;
\exp[-\beta E_{\rm class}^{\rm bare}]\,.
\la{rate-sphaleron-detailed}
\eeq
Here $E_{\rm class}^{\rm bare}$ is the classical energy \ur{Eclass}
of the sphaleron solution with bare (unrenormalized) parameters.
The (dimensionful) quantity $V$ is the physical space
volume arising from the integration over the three translational
zero modes of the sphaleron. Moreover the sphaleron has
three rotational zero modes. The effect of all zero modes
is taken into account in eq. \ra{rate-sphaleron-detailed}
through the factor \cite{McLCar}
$\frac{8\pi^2V}{(g^2\beta)^3}\,N_{\rm tr}^3N_{\rm rot}^3$
with the Jacobians
\begin{eqnarray}
N_{\rm tr}&=&
\Bigl[\fracsm{1}{6\pi}\int d^3{\bf r}\,\Bigl((\bar F_{ik}^a)^2
+(\bar D_k\bar\Phi)^\dagger(\bar D_k\bar\Phi)\Bigr)\Bigr]^\frac{1}{2}
\la{NtNr}\\
N_{\rm rot}&=&\Bigl[\fracsm{1}{6\pi}\int d^3{\bf r}\,\biggl(
\bigl(r^2\delta_{jl}-r_jr_l\bigr)\Bigl(\bar F_{ij}^a\bar F_{il}^a+
(\bar D_j\bar\Phi)^\dagger(\bar D_l\bar\Phi)\Bigr)
-\e_{kij}\Lambda_k^a\bar F_{ij}^a\biggr)\Bigr]^\frac{1}{2}\;,
\nonumber\end{eqnarray}
where $\Lambda$ is the solution of the equation
\beq
\Bigl(-\bar D_l^2+\fracsm{1}{4} \bar\Phi^\dagger\bar\Phi\Bigr)_{ab}
\Lambda^b_k =   \e_{kij}\bar F^a_{ij}\;.
\la{QLam}\eeq

The determinants
$\kappa_{\rm FP}$, $\kappa_{\rm bos}$, $\kappa_{\rm ferm}$
in \eq{rate-sphaleron-detailed}
arise from the Gaussian integration over
fluctuations with periodic time boundary conditions for
bosons
and antiperiodic boundary conditions for fermions.
We can write them in the following form
\begin{eqnarray}
&&\kappa_{\rm FP}=\Bigl[{\det}(-\partial_0^2+\K_{\rm FP})
\Bigr]^\frac{1}{2}
=\prod_n\Bigl(2\sinh(\fracsm{\beta}{2}m_W\omega_n^{\SSS\rm FP})\Bigr)
\la{matsub-FP}\\
&&\kappa_{\rm bos}=\Bigl[{\det}'(-\partial_0^2+\K_{\rm bos})
\Bigr]^\frac{1}{2}
={\prod_n}'\Bigl(2\sinh(\fracsm{\beta}{2}m_W\omega_n)\Bigr)
\la{matsub-bos}\\
&&\kappa_{\rm ferm}=\det(\partial_0+{\cal H}_{\rm ferm})
=\prod_n\Bigl(
2\cosh(\fracsm{\beta}{2} m_W\ee_n)\Bigr)\;,
\la{matsub-ferm}\end{eqnarray}
where $\omega_n^2,\omega_n^{{\SSS\rm FP}\,2}$, and $\ee_n$ are the
eigenvalues of $\K_{\rm bos},\K_{\rm FP}$,
and ${\cal H}_{\rm ferm}\,$, respectively. Our numerics is based on a
calculation of the eigenvalues of discretized versions of these operators;
details can be found in Appendix A and in \cite{rub}.
The primed product in \eq{matsub-bos} means that the zero modes are omitted,
the negative mode $\omega_-^2$, however, contributes to $\kappa_{\rm bos}$.

\section{Renormalization}
\setcounter{equation}{0}

Equation \ra{rate-sphaleron-detailed} for the transition rate contains
ultraviolet divergences arising from the infinite products in
\eqsss{matsub-FP}{matsub-ferm}.
These divergences are removed by the renormalization,
where it is sufficient to renormalize the theory at zero
temperature. Keeping this in mind we split
the right hand side of \eq{rate-sphaleron-detailed}
into  the temperature dependent finite part
and  the divergent part corresponding to zero temperature.
We write
\begin{eqnarray}
\Gamma
&=&\frac{m_W|\omega_-|}
{4\pi\sin(\frac{\beta}{2}m_W|\omega_-|)}\,
\frac{8\pi^2V}{g^6\beta^3}\,(N_{\rm tr}N_{\rm rot})^3\,
\exp\Bigl[-\beta\Bigl(E_{\rm class}^{\rm bare}\la{LA4}\\
&&{}+E_{\rm FP}^{T=0}+E_{\rm FP}^{\rm temp}(T)
+E_{\rm bos}^{T=0}+E_{\rm bos}^{\rm temp}(T)
+E_{\rm ferm}^{T=0}+E_{\rm ferm}^{\rm temp}(T)\Bigr)\Bigr]\;,
\quad\nonumber\end{eqnarray}
where the ultraviolet divergent $T=0$ terms are
\begin{eqnarray}
E_{\rm FP}^{T=0}&=&-\fracsm{1}{2}m_W\Bigl(\sum_n\omega_n^{\SSS\rm FP}-
\sum_n\omega_n^{{\SSS\rm FP}\,0}\Bigr)\;,\nonumber\\
E_{\rm bos}^{T=0}&=&+\fracsm{1}{2}m_W\Bigl({\sum_n}''\omega_n-
\sum_n\omega_n^0\Bigr)\;,\nonumber\\
E_{\rm ferm}^{T=0}
&=&-\fracsm{1}{2}m_W\Bigl(\sum_n|\ee_n|-\sum_n|\ee_n^0|\Bigr)\;.
\la{E-T-0}
\end{eqnarray}
The temperature dependent terms
\begin{eqnarray}
E_{\rm FP}^{\rm temp}(T)&=&-\fracsm{1}{\beta}\Bigl(
\sum_n\ln(1-e^{-\beta m_W\omega_n^{\SSS\rm FP}})
-\sum_n\ln(1-e^{-\beta m_W\omega_n^{{\SSS\rm FP}\,0}})\Bigr)\;,\nonumber\\
E_{\rm bos}^{\rm temp}(T)&=&+\fracsm{1}{\beta}\Bigl(
{\sum_n}''\ln(1-e^{-\beta m_W\omega_n})-
\sum_n\ln(1-e^{-\beta m_W\omega_n^0})\Bigr)\;,\nonumber\\
E_{\rm ferm}^{\rm temp}(T)&=&-\fracsm{1}{\beta}\Bigl(
\sum_n\ln(1+e^{-\beta m_W|\ee_n|})-\sum_n\ln(1+e^{-\beta m_W|\ee_n^0|})\Bigr)
\la{EbEf}\end{eqnarray}
are finite and vanish at zero temperature.
Here ${\sum}''$ stands for the sum over all non-zero,
non-negative modes.

We shall perform the renormalization of the zero-temperature
contribution using the proper-time representation
for the quantities \ra{E-T-0} with the cutoff parameter
$\Lambda$:
\begin{eqnarray}
E_{\rm FP}^{T=0}(\Lambda)&=&
\frac{m_W}{4\sqrt{\pi}}\int_{\Lambda^{-2}}^\infty
\frac{dt}{t^{3/2}}\,\Tr\Bigl(\exp[-t\K_{\rm FP}]
-\exp[-t\K_{\rm FP}^{(0)}]\Bigr)\;,
\la{ptint-FP} \\
E_{\rm ferm}^{T=0}(\Lambda)&=&
\frac{m_W}{4\sqrt{\pi}}\int_{\Lambda^{-2}}^\infty
\frac{dt}{t^{3/2}}\,\Tr\Bigl(\exp[-t {\cal H}_{\rm ferm}^2 ]
-\exp[-t ({\cal H}_{\rm ferm}^{(0)})^2]\Bigr)\;.
\la{ptint-ferm}\end{eqnarray}
In the case of  $E_{\rm bos}^{T=0}$ the proper time
representation is modified to suppress the
negative mode contribution:
\beq
E_{\rm bos}^{T=0}(\Lambda)=
- \frac{m_W}{4\sqrt{\pi}}\int_{\Lambda^{-2}}^\infty
\frac{dt}{t^{3/2}}\,
\left\{
\Tr\Bigl(\exp[-t\K_{\rm bos}]
-\exp[-t\K_{\rm bos}^{(0)}]\Bigr)
+(1-e^{-t\omega_-^2})
\right\}
\la{ptint-bos}
\eeq
In the limit $\Lambda\to\infty$ these integrals diverge since for $t\to 0$
\beq
\Tr\Bigl(\exp[-t\K]-\exp[-t \K^{(0)}]\Bigr)=a t^{-1/2} + b t^{1/2} + \cdots\;,
\la{Tr-small-t-asymptotics}
\eeq
where $\K$ can stand for $\K_{\rm bos},\K_{\rm FP},\K_{\rm ferm}=
{\cal H}_{\rm ferm}^2$. We write the divergent pieces of the r.h.s.~of
\eqsss{ptint-FP}{ptint-bos} as
\begin{eqnarray}
&&E_{\rm FP}^{\rm div}=
\frac{m_W}{4\sqrt{\pi}}
\int_{\Lambda^{-2}}^{\nu_{\rm ren}^{-2}}
\frac{dt}{t^{3/2}}\,
[\Tr\exp(-t\K_{\rm FP})]_{\rm div}\;,
\nonumber\\
&&E_{\rm bos}^{\rm div}= -
\frac{m_W}{4\sqrt{\pi}}
\int_{\Lambda^{-2}}^{\nu_{\rm ren}^{-2}}
\frac{dt}{t^{3/2}}\,
[\Tr\exp(-t\K_{\rm bos})]_{\rm div}\;,
\nonumber\\
&&E_{\rm ferm}^{\rm div}=
\frac{m_W}{4\sqrt{\pi}}
\int_{\Lambda^{-2}}^{\nu_{\rm ren}^{-2}}
\frac{dt}{t^{3/2}}\,
[\Tr\exp(-t \K_{\rm ferm})]_{\rm div}\;,
\la{E-div-def}
\end{eqnarray}
where we define
\beq
[\Tr\exp(-t\K)]_{\rm div} = a t^{-1/2} + b t^{1/2}\;,
\eeq
and $\nu_{\rm ren}m_W$ is the renormalization scale
which is determined below from the value of the Higgs pole mass.
Performing the small $t$ expansion \ra{Tr-small-t-asymptotics}
and integrating over $t$ in \ra{E-div-def} we find
\begin{eqnarray}
\lefteqn{
E^{\rm div}_{\rm FP}(\Lambda)
=\frac{m_W}{64\pi^2}\int d^3{\bf r}\,\biggl[
\fracsm{3}{2}(\nu_{\rm ren}^2-\Lambda^2)
(\bar\Phi^\dagger\bar\Phi-4)}\nonumber\\
&&\qquad\qquad\quad{}+\ln\left(\fracsm{\Lambda^2}{\nu_{\rm ren}^2}\right)
\Bigl(-\fracsm{1}{3}(\bar F^a_{ij})^2
+\fracsm{3}{16}(\bar\Phi^\dagger\bar\Phi-4)^2
+\fracsm{3}{2}(\bar\Phi^\dagger\bar\Phi-4)\Bigr)\biggr]\;,
\nonumber\\
\lefteqn{
E^{\rm div}_{\rm bos}(\Lambda)=-\frac{m_W}{64\pi^2}\int d^3{\bf r}\biggl[
\fracsm{3}{2}(\nu_{\rm ren}^2-\Lambda^2)(4+\nuH^2)(\bar\Phi^\dagger\bar\Phi-4)
+\ln\left(\fracsm{\Lambda^2}{\nu_{\rm ren}^2}\right)
\biggl(\fracsm{41}{6}(\bar F^a_{ij})^2}\nonumber\\
&&{}+6(\bar D_i\bar\Phi)^\dagger(\bar D_i\bar\Phi)
+\fracsm{3}{16}(4+\nuH^2+\nuH^4)(\bar\Phi^\dagger\bar\Phi-4)^2
+\fracsm{3}{4}(8+\nuH^2+\nuH^4)(\bar\Phi^\dagger\bar\Phi-4)\biggr)\biggr]\;,
\nonumber\\
\lefteqn{
E^{\rm div}_{\rm ferm}(\Lambda)=\frac{m_W}{64\pi^2}\sum_F
\int d^3{\bf r}\biggl[
4\nuF^2(\nu_{\rm ren}^2-\Lambda^2)(\bar\Phi^\dagger\bar\Phi-4)+
\ln\left(\fracsm{\Lambda^2}{\nu_{\rm ren}^2}\right)
\biggl(\fracsm{1}{3}(\bar F^a_{ij})^2}\nonumber\\
&&{}+2\nuF^2(\bar D_i\bar\Phi)^\dagger(\bar D_i\bar\Phi)
+\fracsm{1}{2}\nuF^4(\bar\Phi^\dagger\bar\Phi-4)^2
+4\nuF^4(\bar\Phi^\dagger\bar\Phi-4)\biggr)\biggr]\;.
\la{reno3}\end{eqnarray}

The fermionic part in the above formulas
is written for one fermion doublet.
As it was mentioned above, to preserve the spherical symmetry of the Dirac
equation one has to consider the case of equal masses for up and down
fermions. This approximation is justified for doublets
where both fermions are light.
However, in the case of the $(t,b)$ doublet
the approximation $m_b\ll m_W \ll m_t$ is more reasonable.
In this limit the correction
from the top quark is {\it half\/} that of the doublet with both masses
equal to $m_t$ \cite{rub}.
Therefore, in our numerical
estimates we take $9+\frac{3}{2}$ massless fermion doublets
and $\frac{3}{2}$ massive
doublets with mass $m_t$, taking into account that the quark
contribution is enhanced by three colours.

We see that the quadratically ($\Lambda^2$) and logarithmically ($\ln
\Lambda^2$) divergent terms are exactly those entering the classical
energy functional $E_{\rm class}^{\rm bare}$
\ra{Eclass}. Therefore, they can be
combined with the bare constants of the correspondent terms in
the classical energy --- to produce the renormalized constants at the scale
of $\nu_{\rm ren} m_W$. We call the classical energy
with renormalized constants $E_{\rm class}$.

What is left after the renomalization is ultraviolet finite, and one
can safely put the ultraviolet cutoff $\Lambda$ to infinity. We call
these pieces renormalized energies,
\beq
E_{\rm FP}^{\rm ren}=\lim_{\Lambda\to\infty}E_{\rm FP}^{\rm conv}(\Lambda)\,,
\la{flucren}
\eeq
with
\begin{eqnarray}
E_{\rm FP}^{\rm conv}(\Lambda)
&=&\frac{m_W}{4\sqrt{\pi}}
\Biggl[
\int_{\Lambda^{-2}}^\infty
\frac{dt}{t^{3/2}}\,
\Tr\Bigl(\exp[-t\K_{\rm FP}]
-\exp[-t\K_{\rm FP}^{(0)}]\Bigr)\;
\nonumber\\
&& \qquad\qquad\qquad\qquad -\int_{\Lambda^{-2}}^{\nu_{\rm ren}^{-2}}
\frac{dt}{t^{3/2}}\,
[\Tr \exp(-t \K_{\rm FP} )]_{\rm div}\Biggr]\;,\qquad
\la{ptint-FP-ren}
\end{eqnarray}
and similarly for $E_{\rm ferm}^{\rm ren}$ and
$E_{\rm bos}^{\rm ren}$. In the case of $E_{\rm bos}^{\rm ren}$
one should take into account the subtraction of the negative mode
contribution in \ra{ptint-bos} so that
\begin{eqnarray}
E_{\rm bos}^{\rm conv}(\Lambda)&=&
-\frac{m_W}{4\sqrt{\pi}}
\Biggl[
\int_{\Lambda^{-2}}^\infty
\frac{dt}{t^{3/2}}\,
\Bigl[
\Tr\Bigl(\exp[-t\K_{\rm bos}]
-\exp[-t\K_{\rm bos}^{(0)}]\Bigr)+(1-e^{-t\omega_-^2}) \Bigr]
\nonumber\\
&&\qquad\qquad
- \int_{\Lambda^{-2}}^{\nu_{\rm ren}^{-2}}
\frac{dt}{t^{3/2}}\,
[\Tr \exp(-t \K_{\rm bos} )]_{\rm div}\Biggr]\;.
\la{ptint-bos-ren}
\end{eqnarray}

Performing the described renormalization procedure we arrive at the
following expression for the transition rate
\ur{LA4}
\begin{eqnarray}
\gamma&=&\frac{\Gamma}{V} =
\frac{2\pi m_W|\omega_-|(N_{\rm tr}N_{\rm rot})^3}
{g^6\beta^3\sin(\fracsm{\beta}{2}m_W|\omega_-|)}\,
\exp \Bigl[ -\beta \Bigl(E_{\rm class}+E_{\rm FP}^{\rm ren}+
E_{\rm FP}^{\rm temp}(T) \nonumber\\
&&\qquad\qquad
+E_{\rm bos}^{\rm ren}+E_{\rm bos}^{\rm temp}(T)
+E_{\rm ferm}^{\rm ren}+E_{\rm ferm}^{\rm temp}(T)\Bigr)\Bigr]\;.
\la{LA5} \end{eqnarray}

Finally we fix the renormalization point $\nu_{\rm ren}$
in \eqss{E-div-def}{ptint-FP-ren}{ptint-bos-ren} so that the renormalized
parameter $\nu_H$ coincides with the physical pole mass.
In order to obtain the pole mass we have to evaluate the propagator
of the Higgs particle in one-loop order. Its classical part (in Euclidean
space) is given by $G^{-1}(p^2)=p^2+\nu_H^2$ (in units of $m_W^2$),
the fermionic and bosonic one-loop corrections can be written as
$\alpha\,N_c\,\nu_t^4\,
F_{\rm ferm}(\frac{\nu_{\rm ren}^2}{\nu_t^2},\,\frac{p^2}{\nu_t^2})$
and $\alpha\,\nu_{\rm ren}^2\,F_{\rm bos}(\nu_{\rm ren}^2,\,p^2,\,\nu_H^2)$
with some functions $F_{\rm ferm}$ and $F_{\rm bos}$ which
are finite in the limit of infinite $\nu_t$ and $\nu_{\rm ren}$, respectively.

In most parts of our numerical computation we work with a top quark mass
substantially larger than $m_H$. From the classical part of $G^{-1}(p^2)$
we know that $p^2$ is of the order of $-\nu_H^2$, while we find (see below)
$\nu_{\rm ren}^2\sim \nu_t^2$. Therefore, the dominating parts of
the one loop contribution are terms of the order
$\alpha\,N_c \nu_t^4$ and $\alpha\,N_c \nu_t^2$ stemming from the
fermionic correction, so for $m_t$ significantly larger than
$m_H$ it is a good approximation to drop the bosonic contribution
completely (its leading term is only of the order
$\alpha\,\nu_{\rm ren}^2\sim\alpha\,\nu_t^2$) and to expand $F_{\rm ferm}$
in $p^2/\nu_t^2$ up to the order ${\cal O}(p^2/\nu_t^2)$.
Within this approximation we
can also neglect corrections to the pole mass of the W-boson.
Exceptions from this procedure will be treated below.

In order to get the Higgs propagator we put the gauge field to zero and
obtain the relevant (four-dimensional) action:
\beq
S=S_0+S_{\rm loop}\,,\la{act-pole}
\eeq
with
\begin{eqnarray}
S_0&=&\frac{1}{g^2}\int d^4 x\left[\fracsm{1}{2}
       (\partial_\mu\Phi)^\dagger(\partial_\mu\Phi)
      +\fracsm{\nu_H^2}{32}(\Phi^\dagger \Phi)^2-\fracsm{\nu_H^2}{4}
       \Phi^\dagger \Phi\right]\,,
      \\
S_{\rm loop}&=&-\frac{1}{2}\Tr\log D^\dagger D + S_{\rm counter}
      = \frac{1}{2} \int_{\Lambda^{-2}}^\infty \frac{dt}{t}\,\Tr\,
        e^{-tD^\dagger D} -\frac{1}{4\sqrt{\pi}}
        \int_{\Lambda^{-2}}^{\nu_{\rm ren}^{-2}}\frac{dt}{t}\left(
        \frac{a}{t}+b\right)\,,\nonumber
\end{eqnarray}
where we used
\beq
D^\dagger D=-\partial^2+i\Bigl[(i\gamma_0\partial_0+\gamma_i\partial_i)
(M\fracsm{1+\gamma_5}{2} + M^\dagger\fracsm{1-\gamma_5}{2})\Bigr]
+ M^\dagger M\,,
\eeq
and the counterterms
\beq
a=-\frac{N_c\,\nu_t^2}{8\pi^{3/2}}\int d^4x\,\Phi^\dagger\Phi\,,
    \quad\quad\quad
b=\frac{N_c}{64\pi^{3/2}}\int d^4x\left(\nu_t^4(\Phi^\dagger\Phi)^2+
    4\nu_t^2(\partial_\mu\Phi)^\dagger(\partial_\mu\Phi)\right)\,.
\eeq
The vacuum expectation value of the Higgs field in one-loop order
is obtained by setting $\Phi^\dagger\Phi=v^2={\rm const}$
in the action \ur{act-pole}
and minimizing it with respect to $v$. The result up to order $g^2$ is
\beq
v^2=v_0^2+\frac{g^2\,N_c}{2\pi^2\nu_H^2}\left(\nu_t^2\nu_{\rm ren}^2
        -\nu_t^4(1-C-\ln\fracsm{\nu_t^2}{\nu_{\rm ren}^2})\right)\,,
\eeq
where $v_0^2=4$ is the vev.~on the classical level and $C\approx 0.577$
the Euler constant. Choosing the unitary gauge, we can substitute
\beq
\Phi={0\choose v+g\eta}\,.
\eeq
The propagator is then given by
\beq
\frac{\delta^2 S}{\delta\eta(x_1)\,\delta\eta(x_2)}\biggr|_{\eta=0}
   =\int\frac{d^4p}{(2\pi)^4}
    \, e^{ip(x_2-x_1)}\, G^{-1}(p^2)\,,
\eeq
and we obtain up to order $g^2$
\beq
G^{-1}(p^2)=p^2+\nu_H^2+\frac{g^2\,N_c\,\nu_t^4}{8\pi^2}\biggl[
   \fracsm{\nu_{\rm ren}^2}{\nu_t^2}-1-\fracsm{1}{4}\,\fracsm{p^2}{\nu_t^2}
  (C+\ln\fracsm{\nu_t^2}{\nu_{\rm ren}^2})
  -(2+\fracsm{1}{2}\,\fracsm{p^2}{\nu_t^2})f(-\fracsm{p^2}{\nu_t^2})\biggr]
\la{fullprop}
\eeq
with the function
\beq
f(x)\equiv\fracsm{1}{2}\int_0^1 d\beta\,\ln|1-\beta(1-\beta)x|
      =-1+{\rm Re}\left(\sqrt{\frac{x-4}{x}}\,{\rm artanh}\,
     \sqrt{\frac{x}{x-4}}\right)\,.
\eeq
As mentioned before, we expand the propagator up to terms of
${\cal O}(p^2/\nu_t^2)$. We obtain
\beq
G^{-1}(p^2)=p^2+\nu_H^2+\frac{g^2\,N_c\,\nu_t^4}{8\pi^2}\biggl[
   \fracsm{\nu_{\rm ren}^2}{\nu_t^2}-1-\fracsm{1}{4}\,\fracsm{p^2}{\nu_t^2}
  \Bigl(\fracsm{2}{3}+C+\ln\fracsm{\nu_t^2}{\nu_{\rm ren}^2}
  \Bigr)\biggr]\,.
\eeq
Our aim is to fix $\nu_{\rm ren}$ such that the pole mass $\nu_p$,
defined by $G^{-1}(p^2=-\nu_p^2)=0$, coincides with $\nu_H$. Hence we
have to solve the equation
\beq
\nu_{\rm ren}^2=\nu_t^2-\frac{\nu_H^2}{4}\Bigl(\frac{2}{3}+C
   +\ln\frac{\nu_t^2}{\nu_{\rm ren}^2}\Bigr)
\la{nurenfix}\eeq
which determines the renormalization constant $\nu_{\rm ren}$ for given
values of $\nu_t$ and $\nu_H$, $\nu_t\simgt\nu_H$. As anticipated, we
find $\nu_{\rm ren}^2\sim\nu_t^2$.

Analogous equations follow from the full (non-expanded) propagator
\ur{fullprop} and from the propagator which includes both fermionic and
bosonic fluctuations. We have checked that for $\nu_t\simgt\nu_H$ the
results for $\nu_{\rm ren}$ obtained with those equations are very close
to
the solutions of \eq{nurenfix}, so that in this case the restriction to
the dominating parts up to ${\cal O}(\alpha\,N_c \nu_t^2)$ is a very
good
approximation, and for our choice of $\nu_{\rm ren}$ the renormalized
parameter $\nu_H$ corresponds to the Higgs pole mass very accurately.

The situation is slightly worse for the case of very large $\nu_H$
(e.g.~$m_H=350$ GeV). Here
no solution of \eq{nurenfix} can be found, so we choose
$\nu_{\rm ren}$ such that the difference $|\nu_p-\nu_H|$ takes its
minimum.
The deviation, however, is found to be below 10$\%$. Moreover, we
consider such high Higgs masses only for
comparison to our main results, so that this problem is actually
irrelevant.

\section{Thermal renormalization and spectral densities}
\setcounter{equation}{0}

We next consider the temperature dependent terms in the exponent of
\ra{LA4} given by \ra{EbEf}. In order to compute these quantities
we introduce spectral
densities $\vrho^{\cdots}(E)$ for the continuous parts of the spectra of
the operators $\K_{\cdots}$,
such that for any function $f(x)$ that vanishes fast enough for
$x\to\infty$ we have
\beq
\Tr\Bigl(f(\K)-f(\K^{(0)})\Bigr)=
\sum_{\rm dicrete\atop levels}f(\omega^2)
+
\int_0^\infty dE\;\vrho(E)f(E^2)\;.
\la{dens}\eeq
$\K_{\rm bos}$ has $n_{\SSS D}^{\rm bos}=7$ discrete levels
(six zero-modes and one negative mode),
$\K_{\rm ferm}$ has $n_{\SSS D}^{\rm ferm}=1$ discrete zero mode
\cite{rub}, $\K_{\rm FP}$ has $n_{\SSS D}^{\rm FP}=0$ discrete modes,
which by definition are not included in the spectral
densities $\vrho^{\cdots}$.

We rewrite \eq{EbEf} with the help of the spectral densities:
\begin{eqnarray}
E_{\rm bos,FP}^{\rm temp}(T)
&=&\pm\fracsm{1}{\beta}\int_0^\infty dE\,\vrho^{\rm bos,FP}(E)
\,\ln(1 - e^{-\beta m_W E}) \;, \nonumber\\
E_{\rm ferm}^{\rm temp}(T)
&=&-\fracsm{1}{\beta}\int_0^\infty dE\,\vrho^{\rm ferm}(E)
\,\ln(1 + e^{-\beta m_W E})-\fracsm{1}{\beta}\,n_{\SSS D}^{\rm ferm}\ln 2 \;,
\la{E-temp-spectral}
\end{eqnarray}
where the signs are chosen according to those of
$E_{\cdots}^{\rm temp}(T)$ given in \eq{EbEf}.

At high temperatures eq.~\ra{E-temp-spectral}
is dominated by
spectral densities at large energies.
It is shown in Appendix~B that asymptotically
all three spectral densities
$\vrho^{\rm FP}$, $\vrho^{\rm bos}$ and
$\vrho^{\rm ferm}$ approach constant values (after subtraction of
the spectral densities of free operators, which is implied in the
definition \ra{dens}),
$$\lim\limits_{E\to\infty}\vrho(E)=\vrho_\infty\;.$$
Therefore, at high $T$ the quantities $E_{\cdots}^{\rm temp}(T)$
have a $T^2$ behaviour:
\beq
E_{\cdots}^{\rm large}(T)=\pm\fracsm{1}{\beta}\int_0^\infty dE\,
\vrho^{\cdots}_\infty\ln(1\pm e^{-\beta m_W E})
=\pm\frac{\pi^2\vrho^{\cdots}_\infty T^2}{24m_W}\,(-1\pm 3)\;,
\la{Elar}\eeq
whereas the remaining part of $E_{\cdots}^{\rm temp}(T)$ can be
written as (see \eq{intrnd})
\begin{eqnarray}
\lefteqn{\pm\fracsm{1}{\beta}\int_0^\infty dE\,
\Bigl(\vrho^{\rm bos,FP}(E)-\vrho^{\rm bos,FP}_\infty\Bigr)
\,\ln(1 - e^{-\beta m_W E})} \qquad\nonumber\\
&=&\pm\fracsm{1}{\beta}\int_0^\infty dE\,
\Bigl(\vrho^{\rm bos,FP}(E)-\vrho^{\rm bos,FP}_\infty\Bigr)
\,\ln\frac{1 - e^{-\beta m_W E}}{\beta m_W} \mp \fracsm{1}{\beta}
n_{\SSS D}^{\rm bos,FP}\ln(\beta m_W)\qquad \nonumber\\
&\equiv&E^{\rm small}_{\rm bos,FP}\mp\fracsm{1}{\beta}
 n_{\SSS D}^{\rm bos,FP}\ln(\beta m_W)
\la{small-bos}  \\
E_{\rm ferm}^{\rm small}
&=&-\fracsm{1}{\beta}\int_0^\infty dE\,
\Bigl(\vrho^{\rm ferm}(E)-\vrho^{\rm ferm}_\infty\Bigr)
\,\ln(1 + e^{-\beta m_W E}) - \fracsm{1}{\beta}
n_{\SSS D}^{\rm ferm}\ln 2 \;.
\la{small-ferm}
\end{eqnarray}
With these definitions the  $E^{\rm small}_{\cdots}$ are finite in the
high temperature limit (see App.~B).

We notice that at $T={\cal O}(m_W/g)$ the local functionals $E_{\cdots}^{\rm
large}(T)$ can be of the same order as the (renormalized) classical
zero-temperature energy of the sphaleron, therefore in that range of
temperature one has to find a new sphaleron solution
which is a saddle point of the temperature dependent functional
\beq
E_{\rm class}^{\rm ren}(T)= E_{\rm class}+E_{\rm FP}^{\rm large}(T)
+E_{\rm bos}^{\rm large}(T)+E_{\rm ferm}^{\rm large}(T)\; .
\la{Tren}\eeq
Fortunately this functional
has the same form as the original $E_{\rm class}$, but with
temperature dependent parameters.
Therefore its saddle point will be just
a rescaled version of the original sphaleron configuration.

Using the expressions for $\vrho_\infty$
computed in appendix B \eqs{rho0a}{aaa} we arrive at
\begin{eqnarray}
E_{\rm class}^{\rm ren}[A,\Phi;T]&=&\frac{m_W}{g^2}\int
d^3{\bf r}\,\Bigl[\fracsm{1}{4}(F^a_{ij})^2
+\fracsm{1}{2}(D_i\Phi)^\dagger(D_i\Phi)
+\fracsm{1}{32}\nuH^2(\Phi^\dagger\Phi-4)^2\nonumber\\
&&\qquad\qquad\quad{}+\frac{g^2T^2}{32m_W^2}
(\fracsm{2}{3}N_c\nut^2+\nuH^2+3)
(\Phi^\dagger\Phi-4)\Bigr]\nonumber\\
&=&\frac{m_W}{g^2}\int d^3{\bf r}\,\Bigl[\fracsm{1}{4}(F^a_{ij})^2
+\fracsm{1}{2}(D_i\Phi)^\dagger(D_i\Phi)+\fracsm{1}{32}\nuH^2
(\Phi^\dagger\Phi-4q^2)^2\qquad\nonumber\\
&&\qquad\qquad\qquad{}-\fracsm{1}{32}\nuH^2(4-4q^2)^2\Bigr]\nonumber\\
&=&\frac{m_Wq}{g^2}\int d^3{\bf r}\,\Bigl[\fracsm{1}{4}(\tilde F^a_{ij})^2
+\fracsm{1}{2}(\tilde D_i\tilde\Phi)^\dagger
(\tilde D_i\tilde\Phi)+\fracsm{1}{32}\nuH^2
(\tilde\Phi^\dagger\tilde\Phi-4)^2\qquad\nonumber\\
&&\qquad\qquad\qquad{}-\fracsm{1}{32}\nuH^2(4q^{-2}-4)^2\Bigr]
\la{EclassT}\end{eqnarray}
where in the last expression we change the integration variable,
${\bf r}\to q^{-1}{\bf r}$, and use the notations
\beq
\tilde A({\bf r})= q^{-1}A(q^{-1}{\bf r})\;,\qquad\qquad
\tilde\Phi({\bf r})= q^{-1}\Phi(q^{-1}{\bf r})
\la{fieldtild}\eeq
with (compare e.g.~\cite{Kirz})
\begin{eqnarray}
&&q=q(T)=\sqrt{1-\left(\frac{T}{T_c}\right)^2}\;,\nonumber\\
&&T_c=\frac{2\sqrt{2}\,\nuH m_W}{g}\,\Bigl(\fracsm{2}{3}N_c\nut^2
+\nuH^2+3\Bigl)^{-\frac{1}{2}}\;.
\la{qT}\end{eqnarray}

The new temperature dependent sphaleron configuration
$\bar A^q,\bar\Phi^q$ which is the saddle point of
$E_{\rm class}^{\rm ren}(T)$
can be expressed in terms of the old zero temperature solution
$\bar A,\bar\Phi$:
\beq
\bar A^q({\bf r})=q\bar A(q{\bf r})\;,\qquad\qquad
\bar\Phi^q({\bf r})=q\bar\Phi(q{\bf r})\;.
\eeq
Subtracting the vacuum contribution we find
\beq
E_{\rm class}^{\rm ren}[\bar A^q,\bar\Phi^q;T]-E_{\rm class}^{\rm ren}
[\bar A^{q\,(0)},\bar\Phi^{q\,(0)};T]=qE_{\rm class}[\bar A,\bar\Phi]\;.
\la{EclTEcl}
\eeq
If we now replace $\bar A,\bar\Phi$ by $\bar A^q,\bar\Phi^q$
in all parts of our expression for $\gamma$, we find that we
get the old results again but with $m_W$ replaced by $qm_W$ wherever it
appears (except in the definition \ur{qT} of $T_c$).
The vacuum expectation value of the Higgs field becomes
temperature dependent; it is given by $\Phi^{q\,(0)}=2q{0\choose 1}$
and vanishes for $T\to T_c$.

Thus the final result for the transition rate per volume is
\begin{eqnarray}
\gamma&=&
\frac{2\pi qm_W|\omega_-|(N_{\rm tr}N_{\rm rot})^3}
{g^6\beta^3\sin(\fracsm{\beta}{2}qm_W|\omega_-|)}\,
(qm_W\beta)^{n_{\SSS D}^{\rm bos}}\,
\exp\biggl[-\beta q\Bigl(E_{\rm class}+E_{\rm FP}^{\rm ren}
+E_{\rm bos}^{\rm ren}+E_{\rm ferm}^{\rm ren}\Bigr)\nonumber\\
&&\ \qquad\qquad\qquad{}-\beta\Bigl(E_{\rm FP}^{\rm small}(T)
+E_{\rm bos}^{\rm small}(T)
+E_{\rm ferm}^{\rm small}(T)\Bigr){\TS{\ \atop m_W\to qm_W}}\biggr]
\nonumber \\
&=&{\cal F}\exp\biggl[-\beta\Bigl(qE_{\rm class}+E_{\rm FERM}
+E_{\rm BOS}\Bigr)\biggr] \la{gam}
\end{eqnarray}
with the prefactor
\beq
{\cal F}=\frac{2\pi(qm_W)^8\beta^4|\omega_-|(N_{\rm tr}N_{\rm rot})^3}
{g^6\sin(\fracsm{\beta}{2}qm_W|\omega_-|)}\,,
\la{pref}\eeq
and the total fermionic and bosonic one-loop contributions are
\begin{eqnarray}
E_{\rm FERM}&=&
qE_{\rm ferm}^{\rm ren}+E_{\rm ferm}^{\rm small}(T)\Bigr|_{qm_W}\nonumber\\
E_{\rm BOS}&=&
qE_{\rm bos}^{\rm ren}+E_{\rm bos}^{\rm small}(T)\Bigr|_{qm_W}
+qE_{\rm FP}^{\rm ren}+E_{\rm FP}^{\rm small}(T)\Bigr|_{qm_W}\;.
\la{gamE}\end{eqnarray}

\section{Dissipation of the baryon asymmetry}
\setcounter{equation}{0}

In this section we express the erasure of the BAU after the electroweak
phase transition through the transition rate $\gamma(T)$
following the considerations of
\cite{Arnold} (see also \cite{Boch}).

Transitions over the sphaleron barrier change the baryon and lepton
number by
$\Delta B=\Delta L=N_g\,\Delta\NC$, where $N_g=3$ is the number
of generations.

If there were neither baryons nor leptons initially,
the transitions increasing $\Ncs$ would be compensated by
processes decreasing $\Ncs$ so that
creation and annihilation of baryons
would cancel each other. The situation becomes different, however, if
we assume the initial existence of baryons and leptons.
These particles could have been created long time before the
electroweak phase transtion in the age of GUT or maybe during the phase
transition. In this case the chemical potentials $\mu$ of the baryons
and leptons
are non-zero which leads to an additional term $\mu\NC$ in the classical
energy functional.
In accordance with the Le Ch\^atelier
principle the transitions will favor the wash-out of any particle or
antiparticle excess.

Our analysis will be restricted to the usual electroweak theory
where  $B-L$ is conserved.
Therefore, the study of the erasure of the BAU in the
context of sphaleron transitions makes sense only under
the assumption of the initial condition $B=L$.

A sphaleron transition with $\Delta\NC=1$ creates one particle per
fermion doublet. We introduce chemical potentials $\mu^Q_i$
($i=1,\ldots,9$) for the quark doublets and $\mu^L_i$ ($i=1,\ldots,3$)
for the lepton doublets. Then each transition increases the energy by
$\Delta\NC\left(\sum_{i=1}^9\mu^Q_i+\sum_{i=1}^3\mu^L_i\right)$, i.e.~the
classical energy functional \ur{Eclass} has to be replaced by
\cite{Redlich,DiPet}
\beq
E_{\rm class}\to E_{\rm class}+
\NC\left(\sum_{i=1}^9\mu^Q_i+\sum_{i=1}^3\mu^L_i\right)\, .
\la{enrepl}\eeq
This $\mu$ contribution leads to the asymmetry in the Langer--Affleck
formula \ra{gam} with respect to transitions increasing and decreasing
$\Ncs$.  We have to set $\NC=+\frac{1}{2}$ for transitions which increase
the fermion number and $\NC=-\frac{1}{2}$ for transitions which decrease
it.  Since the baryon and lepton densities considered are very small we can
restrict the transition rate to terms linear in $\mu$
\begin{eqnarray}
\gamma_+&=&\gamma\left[1-\frac{\beta}{2}
\left(\sum_{i=1}^9\mu^Q_i+\sum_{i=1}^3\mu^L_i\right)\right]\, ,
\nonumber \\
\gamma_-&=&\gamma\left[1+\frac{\beta}{2}
\left(\sum_{i=1}^9\mu^Q_i+\sum_{i=1}^3\mu^L_i\right)\right]\, ,
\la{partrate}
\end{eqnarray}
so that the baryon and lepton number dissipation, given by the
difference of these rates, reads:
\beq
\frac{dB}{dt}=\frac{dL}{dt}=-\gamma(T)\,VN_g\,\beta
\left(\sum_{i=1}^9\mu^Q_i+\sum_{i=1}^3\mu^L_i\right)\, .
\la{barrate}\eeq

Now we have to express the chemical potentials through the particle
numbers in order to get a differential equation for the baryon number
decrease. For small $\mu$
standard Fermi--Dirac statistics yields the relation
\beq
N(\mu)=\frac{fV}{6\beta^2}\,I(\beta m)\,\mu\, ,
\la{fermistat}\eeq
where $N$ is the number of fermions, $f$ the number of degrees of
freedom, $m$ the mass of the fermions and $I(a)$ is given by
\beq
I(a)=\frac{12}{\pi^2}\int_a^\infty\frac{dx}{1+e^x}\,
\frac{x^2-\frac{1}{2}\,a^2}{\sqrt{x^2-a^2}}\, ,
\la{massint}\eeq
with the properties $I(0)=1$, $I(\infty)=0$. A lepton doublet has
three degrees of freedom (two for the lepton and one for the
neutrino) while there are four for a quark doublet. The masses of
the leptons and the light quarks are much smaller than the critical
temperature $T_c=1/\beta_c$, i.e.~$\beta_c m\ll 1$ so that
$I(\beta_c m)\approx 1$. We apply this approximation also in the
case of the top quark since it has almost no influence on
the result for the baryon number wash-out.
Hence we obtain for the lepton and the light quark doublets:
\beq
\mu_i^L=\frac{2\beta^2}{V}\,L_i\, ,\quad\quad\quad
\mu_i^Q=\frac{3\beta^2}{2V}\,Q_i\, , \la{chempot}
\eeq
where $L_i$, $Q_i$ are the numbers of leptons and quarks of a fixed
doublet $i$. Substituting
\eq{chempot} into \eq{barrate} and using $L=B=Q/3$ we obtain:
\beq
\frac{dB}{dt}=-\,\gamma(T)\,VN_g\,\beta\left\{\frac{3\beta^2}{2V}\,Q
    +\frac{2\beta^2}{V}\, L\right\}
   =-\fracsm{13}{2}\,\gamma(T)\,N_g\,\beta^3B\;.
 \la{subrate}\eeq
Standard cosmology gives a relation between time and temperature
\cite{Weinberg,Olive}:
\beq
t=\sqrt{\frac{45}{16\pi^3 N(T)}}\,m_P T^{-2}=C\,T^{-2}\, ,
\eeq
where $N(T)$ is some number related to the number of degrees of freedom
of the thermalized particles at the temperature $T$ (for our range of
temperature it is usually $\frac{381}{4}$), $m_P=1.5\cdot
10^{17}\,m_W$ is the Planck mass and hence the constant $C$ is given by
$C\approx 5\cdot 10^{15}\, m_W$.
Substitution yields
\beq
\frac{1}{B}\,\frac{dB}{dT}=13N_g\,C\,\frac{\gamma(T)}{T^6}\, ,
\eeq
which can be integrated to
\begin{eqnarray}
B(T)&=&B(T_c)\,\exp\left\{-13N_g
      \,C\int_T^{T_c}\frac{\gamma(T)}{T^6}\,dT\right\}\nonumber \\
      &=&B(T_c)\,\exp
     \left\{\frac{-13N_g\,C}{T_c^5}\int_0^{q(T)}\frac{q\,\gamma(q)}
       {(1-q^2)^{7/2}}\,dq\right\}\;,
\la{Bquot}
\end{eqnarray}
with $q(T)=\sqrt{1-T^2/T_c^2}$. This is our final result; it
describes how the erasure of the BAU, measured by $B_0/B_{T_c}$, can be
obtained by an integration of $\gamma(T)$ over the temperature. In the
next two sections we present numerical results of this ratio from which we
deduce an upper bound on the Higgs mass.

\section{Numerical Results}
\setcounter{equation}{0}

In this section we present the results of our numerical calculations.
We take $m_W=83$ GeV, $g=0.67$ which is the physical value of
the coupling constant and vary the
top quark mass in the range 150 to 200 GeV, i.e.~around its recently
stated value of 174 GeV \cite{topmass}.
The only unknown physical parameter left is the Higgs mass $m_H$.
We discussed the fermion loop contribution already in \cite{rub} so in
this work we will focus our attention on the evaluation of the bosonic
loops.

As we explained, all our results are obtained by a
diagonalization of the boson fluctuation matrix ${\cal K}_{\rm bos}$
and the Faddeev--Popov matrix ${\cal K}_{\rm FP}$ \ur{Kfluc} in a discretized
basis. The spectrum of the matrices must not depend on the choice of
the gauge for the classical sphaleron fields. Since we evaluated the
matrix for an arbitrary gauge with non-vanishing $C$-field (see
\eq{hedgemat}), it is possible to check
the invariance of the spectrum under gauge transformations \eq{HeHotr}.
We find that the eigenstates are indeed invariant under gauge
transformations if their energy is less than about $0.8\,P_{\rm max}$
where
$P_{\rm max}$ is the numerical parameter which restricts the momentum
of the eigenstates and renders the basis finite to allow a numerical
diagonalization (see \eq{discr}).
Eigenstates with energies close to $P_{\rm max}$ can be
gauge dependent which is due to the finite numerical
box and should not be encountered in our calculations.
Hence we always have to
choose $P_{\rm max}$ large enough so that all eigenstates which enter
the calculations have energies less than $0.8\,P_{\rm max}$ and, of
course, no result changes if $P_{\rm max}$ is further increased.

Another check of the spectrum consists in an investigation of the
negative and zero modes. The negative mode appears in the
grand-spin $K=0$ sector of the fluctuation matrix ${\cal K}_{\rm bos}$.
We
checked that its energy is gauge invariant
and independent of the box parameters $P_{\rm max}$ and $R$. Moreover
our
results agree with the ones obtained in \cite{AKY1,Brihaye} where the
negative mode has previously been calculated. The zero modes can be
identified in the $K=1$ sector. Due to the spherical symmetry in this
sector each state is (2K+1=3)-fold degenerate so that we find two
threefold degenerate states with zero eigenvalue. Numerically the
modulus of the eigenvalues was found to be below $10^{-3}$ which shows
that the diagonalization reproduces the zero modes with excellent
accuracy. The eigenfunctions of the zero modes can be evaluated
analytically in terms of the sphaleron background fields \cite{McLCar}.
We compared these functions with those which we obtained as zero mode
eigenfunctions in the diagonalization and again found a very good
agreement.

Beside these investigations of specific eigenstates we checked a
property of the spectrum as a whole. For low values of the proper time
parameter $t$ we consider the expansion
\begin{eqnarray}
\sum_n\left(e^{-t\omega_n^2}-e^{-t{(\omega_n^0})^2}\right)
&=&\Tr\Bigl(\exp[-t{\cal K}_{\rm bos}]-\exp[-t{\cal K}_{\rm bos}^{(0)}]\Bigr)
\nonumber\\
&=&a_{\rm bos}t^{-1/2}+b_{\rm bos}t^{1/2}+c_{\rm bos}t^{3/2}+\cdots
\la{heatk}\end{eqnarray}
where the coefficients are given in \eqsss{aaa}{cbos}.
Taking $m_H=m_W$, in Fig.~1 we compare the exact result (solid line) for
the trace
$\Tr\Bigl(\exp[-t{\cal K}_{\rm bos}]
-\exp[-t{\cal K}_{\rm bos}^{(0)}]\Bigr)$
(l.h.s.~of \eq{heatk}) with several
approximations (dashed and dotted lines), given by the first,
the first two and all three terms of the r.h.s.~of \eq{heatk}.
For low values of $t$ we obtain excellent agreement between the
numerical result and the approximations, as it should be. For large
values of $t$ the comparison does not provide a check of the numerical
treatment. Here the approximations behave as some power law
of $t^{1/2}$ while
the exact result is dominated by the contribution
of the negative and zero modes, given by $e^{t|\omega_-^2|}+6$,
which is also plotted in Fig.~1 (dashed line).

Now that we have checked the reliability of the spectrum we can use it
to calculate the desired quantities. These are the renormalized
non-thermal contributions $E^{\rm ren}_{\rm bos,FP}$ and the
temperature dependent parts $E^{\rm small}_{\rm bos,FP}(T)$
associated with the Bose--Einstein distribution factor.
Both quantities have to be evaluated
for the fluctuation operator ${\cal K}_{\rm bos}$ and the Faddeev--Popov
operator ${\cal K}_{\rm FP}$.

To obtain the renormalized value of the non-thermal parts
$E_{\rm bos,FP}^{\rm ren}$ we have to evaluate
$E_{\rm bos,FP}^{\rm conv}(\Lambda)$ in the
limit of infinite proper time cutoff ($\Lambda\to\infty$)
(see \eq{flucren}).
Numerically, however, we
always have to work with a finite $\Lambda$ to ensure the finiteness of
the basis. For this reason the numerical parameters $R$ (box size) and
$P_{\rm max}$ (maximum momentum) also have to be finite. In order to
obtain
the limit of infinite parameters we proceed as for the fermion
non-thermal energy in \cite{rub}:
First we fix $\Lambda$ and take $R$ and $P_{\rm max}$
large enough so that their further increasement would not change the
result any more. This procedure is repeated with larger values of the
cutoff $\Lambda$ until we can determine the
limit $E^{\rm conv}(\Lambda=\infty)=E^{\rm ren}$. We illustrate this
method in Tabs.~1 and 2 at the example $m_H=m_W$. The renormalization
scale $\nu_{\rm ren}$ is fixed according to \eq{nurenfix}. With
$m_t=174$ we obtain $\nu_{\rm ren}=2.02$. In Tab.~1 we show
results of $E_{\rm bos}^{\rm conv}(\Lambda)$ for fixed $\Lambda=4$ and
various values of $R$ and $P_{\rm max}$. We find that for
$R=12$ and $P_{\rm max}=16=4\,\Lambda$ the continuum limit
$E_{\rm bos}^{\rm conv}(\Lambda=4)=-6.28\,m_W$ is reached with
an accuracy of better than $0.2\%$. With the same method we obtain results
for other values of $\Lambda$ which are presented in Tab.~2. For large
$\Lambda$ the law $E_{\rm bos}^{\rm conv}(\Lambda)=a+b/\Lambda^2$ is
satisfied for the fit $a=-5.95\,m_W$ and $b=-5.7\,m_W$. Thus from the
data in Tab.~2 we can extrapolate the result for infinite cutoff
$\Lambda$ and obtain $E_{\rm bos}^{\rm ren}=a=-5.95\,m_W$.
Considering the possible error of the values for fixed $\Lambda$ and
the error of the extrapolation process we estimate an accuracy of better than
$2\%$ for the final result.
For other Higgs masses the deviations can be slightly bigger but in
general the numerical error for the non-thermal energy
should be well below $5\%$.

Table 3 shows results for the renormalized non-thermal energy
for different Higgs masses including the contribution from the
Faddeev--Popov operator. The renormalization scale $\nu_{\rm ren}$,
fixed by \eq{nurenfix}, is included in Tab.~3. As already mentioned
in Sect.~4, for $m_H=350$ GeV \eq{nurenfix} has no solution so that
instead we choose $\nu_{\rm ren}$ to minimize the difference between
$m_H$ and the pole Higgs mass, which for this reason may deviate
from 350 GeV by some value less than 10$\%$.
Moreover we give values for the
classical sphaleron energy and for the fermion non-thermal energy.
The latter was
calculated for $\frac{3}{2}$ heavy doublets with top quark masses
between 150 and 200 GeV and $9+\frac{3}{2}$ massless doublets.
We find that
both the fermion and the boson non-thermal energy are significantly
lower than the classical energy which is in accordance with the fact that,
generally speaking, loop contributions are suppressed
by a factor $\alpha$ relative to the tree contribution. Actually
after the renormalization the non-thermal energy of the
boson fluctuations about the sphaleron is small and negative while
that of the fermions is larger and positive.
However, we show below that the thermal part dominates the boson
fluctuations and the sum of both parts has the same sign as the
classical energy.

The behavior of the non-thermal energies for low $m_H$ and large
$m_t$ can be described by simple scaling laws. For $m_t/m_W>1$ the
aggregate energy density of the Dirac sea is
dominated by the square loop diagram in the external Higgs field and
hence is proportional to $N_c(h\Phi)^4\ln(h\Phi/\nu_{\rm ren})$
where $h$ is the
Yukawa coupling and $\Phi$ the Higgs field of the sphaleron. To obtain
the energy we have to integrate this value over the space where the
Higgs field differs from its vacuum expectation value, i.e.~over the
spread of the sphaleron.
For small Higgs masses $m_H/m_W\ll 1$ the size of the sphaleron
fields roughly scales as $m_H^{-1}$ since the asymtotic behavior for
large radial distance $r$ is dominated by the term $e^{-(m_H/m_W)\,r}$.
Therefore all spatial integrals and hence all matrix elements of the
fluctuation matrices scale as $m_H^{-3}$. Thus for $m_H<m_W<m_t$
the dependence of the zero temperature energies on $m_t$ and $m_H$ is
roughly given by
\begin{eqnarray}
E_{\rm ferm}^{\rm ren} &\propto&
 + N_c\, \frac{m_t^4\ln(m_t/\nu_{\rm ren}m_W)}{m_H^3}\,, \la{ferzero} \\
E_{\rm bos}^{\rm ren}+E_{\rm FP}^{\rm ren} &\propto&
-\frac{m_W^4}{m_H^3}\,. \la{boszero}
\end{eqnarray}
These scaling laws explain the strong increase of the fluctuations for
small $m_H$ and large $m_t$, which is in correspondence to our
numerical results. As mentioned above, we also found numerically that
the boson and fermion non-thermal energies differ in sign.

One finds a strong increase for small $m_H$ and large $m_t$ also
for the thermal parts of the fluctuations, but here both fermionic
and bosonic contributions are positive.
For not too small temperature $T$ the boson fluctuation energy is
dominated by its positive thermal contribution so that the sum of the
non-thermal and thermal parts is positive for both the boson and the
fermion fluctuation.
For small $m_H$ and large $m_t$ these sums are large so that they
provide a
strong suppression of the sphaleron transition rate. If $m_H$ is
small enough, this suppression prevents the erasure of the BAU. Thus,
we see that the condition that the BAU should survive sets an upper
limit on the Higgs mass.

In order to obtain a quantitative result for this upper bound
we still have to evaluate the thermal parts $E^{\rm small}\Bigr|_{qm_W}$
(see \eqs{small-bos}{small-ferm}).
To simplify notations we will drop the subscript $qm_W$ in what follows.
In principle we could evaluate them by a summation over
the whole spectrum; numerically, however, it is preferable to sum
only over eigenstates with low or medium energy and to use the expansion
\beq
\vrho(E)=\vrho_\infty+\vrho_2\,\frac{1}{E^2}+\cdots
\la{HEexp}\eeq
(see \eq{rhoasym}) for the high energy part of the
spectrum. To this end we take a smooth function $F(E)$
with the properties
$F(E)=1$ for $E<E_a-E_b$, $F(E)=0$ for $E>E_a+E_b$ and $0<F(E)<1$ for
$E_a-E_b<E<E_a+E_b$. Here $E_a$ and $E_b$ are fixed numerical energy
cutoffs, usually we take $E_b=E_a/2$ so that $E_a$ is left as the only
parameter. We calculate the thermal energies as follows:
\begin{eqnarray}
E^{\rm small}_{\rm bos,FP}(T)&=&\pm\frac{1}{\beta}\biggl[
{\sum_n}''F(\omega_n)\,\ln\frac{1-e^{-\beta q m_W \omega_n}}{\beta q  m_W}
 -\sum_n F(\omega_n^0)\,\ln\frac{1-e^{-\beta q m_W \omega_n^0}}{\beta q m_W}
\nonumber \\
&&\qquad\ {}-\vrho_\infty\int_0^\infty \,dE \,F(E)\,\ln
   \frac{1-e^{-\beta q m_W E}}{\beta q m_W}\nonumber \\
&&\qquad\ {}+\vrho_2\int_0^\infty \,dE \,\frac{1-F(E)}{E^2}\,\ln
   \frac{1-e^{-\beta q m_W E}}{\beta q m_W}\biggr]  \la{numkapp}
\end{eqnarray}
The integrals in \eq{numkapp} can easily be evaluated numerically, in
the sum only states with $\omega_n<E_a+E_b$ appear. Now we have to check
that numerically $E^{\rm small}$ is independent of the parameter
$E_a$. In Table 4 we
show results of $\beta E_{\rm bos}^{\rm small}(T)$ in the high
temperature limit (see \eq{boshT})
for $m_H=m_W$ and several values of $E_a$. We also give results for the
contributions (the sum and the integrals of \eq{numkapp}) separately.
We find that in the interval
$3<E_a<8$ both the sum and the integrals in \eq{numkapp}
drastically depend on $E_a$ but the result for
$\beta E_{\rm bos}^{\rm small}$ varies only by about $2\%$. For
smaller values of $E_a$ the expansion \eq{HEexp} is not good any more
and for larger values of $E_a$ the numerical accuracy of the spectrum
decreases due to contributions of states with very large grand spin $K$.
For other Higgs masses the variation of $E_{\rm bos}^{\rm small}(T)$
with $E_a$ can be about $5\%$.
Hence for our calculations we choose $4\le E_a \le 6$ and obtain
$E^{\rm small}(T)$ with an accuracy of usually better than $5\%$.

We are now going to compare our results to the ones obtained by
Carson et.~al.~\cite{McLetal} and Baacke et.~al.~\cite{Baacke}.
There the expression
\beq
\ln\kappa\ldef -\beta E_{\rm bos}^{\rm small}(T)
-\beta E_{\rm FP}^{\rm small}(T)-6\ln2-\ln|\omega_-|
\eeq
was evaluated in the high temperature limit $T\to T_c$. In Fig.~2
we show the results of our work as well as those of
\cite{McLetal,Baacke} as a function of the Higgs mass $m_H$. Our data
are between those of \cite{McLetal} and \cite{Baacke}, they agree with
the ones of \cite{Baacke} up to $10\%$. Apart
from numerical uncertainties one reason for the difference could be the
renormalization scheme. We have performed the renormalization at zero
temperature strictly, as it is usually done. This corresponds to a
subtraction of the first term ($\vrho_\infty$) in
the high energy expansion (see \eq{HEexp}) which is also
the first term of the tadpole expansion \cite{Dolan}.
In \cite{Baacke}, however, all tadpole graphs except the term linear
in $T$ have been removed.
The difference is then due to the higher order terms which are small for
high $T$ but numerically not completely negligible. There is a larger
deviation from the results of \cite{McLetal}, only a qualitative
agreement for the low $m_H$ behavior is found.

Since the evaluation of the fluctuation determinants
\eqsss{matsub-FP}{matsub-ferm} is a rather involved task one seeks for
a good approximation procedure which is easy to handle. Before an exact
calculation was performed, Carson and McLerran \cite{McLCar} applied an
approximation technique by Diakonov, Petrov, and Yung (DPY) \cite{DPY}
to this problem. Later they calculated the boson fluctuation determinant
exactly in the high temperature limit \cite{McLetal} and found that the
exact and the approximative result deviate by several orders of magnitude.
In this section we revisit the DPY method and explain how it can be
used to get the the fluctuation determinants to a reasonable accuracy.

Following \cite{McLCar} we consider here the high temperature limit
of the boson fluctuation determinant which can
be written as an integral over spectral densities:

\beq
\ln\chi_{\rm bos} \equiv -\beta_c E^{\rm small}_{\rm bos}(T_c)
  = -\fracsm{1}{2}\int_0^\infty dE\,
\Bigl(\vrho^{\rm bos}(E)-\vrho^{\rm bos}_\infty\Bigr)\,\ln(E^2)\,.
\la{spectr}\eeq
where $\vrho^{\rm bos}(E)$ is the spectral density of the boson
operator, with the six zero and one negative mode subtracted. Using the
identity
\beq
\ln\alpha=\int_0^\infty \frac{dt}{t}\,\Bigl(e^{-t}-e^{-t\alpha}\Bigr)
\eeq
we arrive at the proper-time representation:
\begin{eqnarray}
\ln\chi_{\rm bos} &=&
 -\fracsm{1}{2}\int_0^\infty dE\,
\Bigl(\vrho^{\rm bos}(E)-\vrho^{\rm bos}_\infty\Bigr)\,
\int_0^\infty \frac{dt}{t}\,\Bigl(e^{-t}-e^{-tE^2}\Bigr)\nonumber \\
&=&\fracsm{1}{2}\int_0^\infty \frac{dt}{t}\,\biggl\{
\Bigl(\Tr\exp[-t\K_{\rm bos}]-6-\exp(t|\omega_-|^2)
   -\Tr\exp[-t\K_{\rm bos}^{(0)}]\Bigr) \nonumber \\
  && \quad\quad\quad\quad  +7e^{-t}
  -\frac{\vrho^{\rm bos}_\infty}{2}\sqrt{\frac{\pi}{t}}\biggr\}\;
   \la{propertime}\end{eqnarray}
where $\vrho^{\rm bos}_\infty= 2a/\sqrt\pi$ (see \eq{rho0a}).

The idea of the DPY method is as follows \cite{DPY}. The behavior
of the integrand at small $t$ can be established from the semiclassical
expansion of the "heat kernel", $\Tr \exp(-t\K)$, see
\eqs{Tr-small-t-asymptotics}{heatk} and \ur{Fexp}.
Its behavior at
large $t$ is governed by negative and zero modes. Therefore,
knowing the behavior of the integrand both at small and large $t$,
one can approximate the integral of \eq{propertime}
as a sum of small- and large-$t$
contributions, separated by some parameter $t_0$,
\beq
\ln\chi_{\rm bos}=\int_0^\infty dt\,f(t)\approx\int_0^{t_0} dt\,f_{\rm low}(t)
+\int_{t_0}^\infty dt\,f_{\rm high}(t)\equiv\ln\tilde\chi_{\rm bos},
\la{separation}\eeq
where the separation parameter should be found from the requirement
that the sum of the two terms in this equation is stable in $t_0$.
Actually it means that $t_0$ is a point where the small- and large-$t$
approximations to the true integrand cross. If that does not
happen the method fails. The more terms one knows from both sides,
the better is the accuracy of the method.

Using the heat kernel expansion for $\Tr\exp(-t\K)$, we find the
approximation for small $t$
\beq
f_{\rm low}(t)=\frac{1}{2}\biggl(bt^{-\frac{1}{2}}+ct^{\frac{1}{2}}
-(7+|\omega_-|^2)+\frac{t}{2}(7-|\omega_-|^4)\biggr)\,.
\la{smexp}\eeq
At large $t$ it behaves as
\beq
f_{\rm high}(t)=\frac{1}{2}\biggl(\frac{7e^{-t}}{t}-at^{-\frac{3}{2}}
\biggr)\,.
\la{laexp}\eeq
Knowing the coefficients $a$, $b$, $c$, and $|\omega_-|$, one can estimate
the fluctuation determinant, \eq{propertime}, using \eq{separation}. The
result for three different values of the Higgs mass is presented in
Table 5, together with the exact value of $\ln\chi_{\rm bos}$. One observes
that the accuracy is at the level of 10 to 15$\%$, but that is a price one
has to pay if one wishes to avoid a laborous computation of the exact
spectrum.

\section{The upper bound for the Higgs mass}
\setcounter{equation}{0}

The main issue of our work is the calculation of the sphaleron
transition rate $\gamma=\Gamma/V$ according to the Langer--Affleck formula
\eqs{LA1}{gam} including the classical Boltzmann factor, the fermionic and
bosonic
one-loop contributions and the Jacobian prefactors. We stress again that
our calculation is not based on the high temperature limit but was
done for arbitrary values of $T$. Therefore, we can compute the rate for
the whole temperature range between zero and the critical temperature
$T_c$ and perform the integration
over $T$ (see \eq{Bquot}) to obtain the ratio $B_0/B_{T_c}$. In Fig.~3
we present the contributions $-\beta qE_{\rm class}+\ln{\cal F}$
(classical part), $-\beta E_{\rm FERM}$ (fermion loop),
and $-\beta E_{\rm BOS}$ (boson loop) of
$\ln\gamma$ (according to \eqsss{gam}{gamE}) for $m_H=66$ GeV
and $m_H=100$ GeV.
It is convenient to take the parameter $q=\sqrt{1-T^2/T_c^2}$ as
independent variable rather than the temperature $T$ itself.

Qualitatively both pictures of Fig.~3 show the same behaviour, but we
find significant quantitative differences. At low temperatures (large $q$)
the main contribution to the classical part is the Boltzmann exponent
$-\beta qE_{\rm class}$ which decreases with $q$ roughly linearly.
For $T\to T_c$ ($q\to 0$) the Jacobian prefactor $\ln{\cal F}
\sim 7\ln q(T) \to -\infty$ dominates the classical part.
Hence we find a maximum of the classical contribution to the
transition rate at about $q\sim 0.1$. For large and medium $q$
the suppression from the fermion loop contribution can be also
rather large (in the
case $m_H=66$ GeV it becomes almost as large as the classical one)
but it tends to zero in the high $T$ limit $q\to 0$. The bosonic
contribution is generally rather small and almost constant over
the plotted range of temperatures. In the high temperature limit
it does not disappear but it tends to some finite value.
Therefore we see that in this limit which was assumed in
previous works \cite{Arnold,Boch,McLCar,McLetal,Baacke}
the boson one-loop contribution
is indeed the most important one while fermions decouple.
Adding the loop contributions to the classical part, we obtain the
total rate which has the same shape as the classical curve, in particular
it also has a maximum. Hence if there was any significant baryon number
violation after the electroweak phase transition,
it must have happened in a short period
around this maximum (remember that in Fig.~3 the logarithm $\ln\gamma$
is plotted while $\gamma$ itself enters the integral of \eq{Bquot}).
We find that
the piece of the curve for the total rate which is marked by a solid line
contributes about 99$\%$ to the ratio $\log_{10}(B_0/B_{T_c})$ which measures
the washout of the BAU.

Both the position of this washout area and the value of the maximum
are strongly influenced by the loop corrections, especially we note
that in this region the fermionic contribution which was neglected
in previous works is quite
essential. Below we investigate the effect of the fermions
quantitatively by computing the ratio $B_0/B_{T_c}$ with and without
fermion loop corrections and confirm the significance of the fermions.

Comparing the two plots of Fig.~3, we find that the loop contributions
are strong for low $m_H$ so that in this case the rate is suppressed,
while for large $m_H$ the fluctuations are rather weak.
This is also documented in Fig.~4, where we have plotted the total
transition rate for various $m_H$ and $m_t$. In accordance to the
scaling laws \eqs{ferzero}{boszero} we find a strong suppression
of the transition rate $\gamma$ for small $m_H$ and large $m_t$
and a weak suppression for large $m_H$. This results finally for
physically relevant $m_t$ in a small transition rate for small $m_H$
and a large transition rate for large $m_H$. If the maximum value of
$\gamma$ is small enough,
the baryon number violating processes have happened so rarely that
they have not affected the BAU. On the other hand a large transition
rate means that the sphaleron transitions must have eliminated the baryon
asymmetry. Thus, if we fix $m_t$ we obtain an upper bound for $m_H$
below which the asymmetry is conserved and above which we expect
a dissipation of the baryon number. This is how we deduce our upper
bound from the condition that the BAU should survive the age of
sphaleron transitions.

Before we evaluate this conclusion quantitatively a comment on the
limits of the approach is in order. Our calculation is based on the
assumption that the
Langer--Affleck formula is valid and our restriction to one-loop
contributions is justified, i.e.~that
higher order corrections can be neglected. Both assumptions are
reasonable for temperatures not too close to $T_c$
and if the fluctuations on the one-loop level are small compared
to the classical part. The physics of the phase transition and in
its direct vicinity, where perturbation expansion breaks down, is
complicated and not well understood yet so it is difficult to decide
at what temperature the framework of our calculation becomes
inapplicable and what in this
case could be a more adequate description.
However, we can estimate the reliability
of our model by checking if, first, in the washout region the
loop contributions are not too big compared to the classical part,
and second, if the onset of the sphaleron transitions, i.e.~the
left margin of the interval marked by a solid line in the corresponding
curve of Fig.~4, is not too close to
the critical temperature. Figs.~3 and 4 show that both
conditions are well fulfilled for $m_H\simgt 100$ GeV but not so good
for smaller $m_H$; for $m_H\simlt 60$ GeV the fluctuations are rather
large so that the model on the one-loop level is probably not
reliable. The conclusion of this
restriction on the applicability of our technical framework will be
drawn later.

Furthermore, we assume that, before the transitions start, there is the same
number of baryons and leptons in the Universe, i.e.~$B-L=0$. In the
standard model $B-L$ is strictly conserved so that this condition will
not change during the period of the transitions. If there were a
primordial excess of e.g.~antileptons, created by unknown forces which
violate $B-L$ before the electroweak phase transition, then the
sphaleron transitions would increase rather than decrease the BAU.

Let us also briefly comment on the connection between our critical
temperature $T_c$ defined in \eq{qT} and the electroweak phase
transition. As a consequence of the thermal renormalization the
vacuum expectation value of $\Phi$ becomes $T$ dependent and vanishes
for $T\to T_c$. This looks like the behavior of fields at a second order
phase transition. However, in order to obtain the true temperature and
nature of the phase transition, one would have to include other terms,
e.g.~a term of the order of $T\Phi^3$ into the potential. Since there is
no consistent way how to perform calculations near the phase transition,
where perturbation theory is not applicable, we decided to take only the
numerically by far dominating term of the order $T^2\Phi^2$ explicitely
into the potential; other terms, like the $T\Phi^3$ one, are considered
in the quantum correction $E^{\rm small}$. Thus our critical temperature
should be seen as a mere definition which needs not necessarily coincide
with the temperature of the true transition, neither do we imply that
the phase transition is of second order.

Knowing the transition rate $\gamma$ as a function of $q$ for a fixed
Higgs mass $m_H$ it is now
possible to perform the integration in \eq{Bquot} numerically.
The result for
the ratio $\log_{10}(B_0/B_{T_c})$ is plotted in Fig.~5
for $m_t=$150, 174, and
200 GeV. For comparison, we also performed the calculation without
considering fermions. In this case we did not use \eq{nurenfix} to
fix the renormalization scale, but a corresponding equation which
follows from the Higgs propagator with boson fluctuations included
instead of fermions. Here we obtain $\nu_{\rm ren}\sim 1$.

All the curves start at zero for small Higgs masses which means that
the BAU survives completely. If we increase $m_H$, the fluctuations
become weaker so that the transition rate increases. Hence the ratio
$B_0/B_{T_c}$ suddenly begins to fall, and within a short interval it drops
by 20 orders of magnitude. The bigger the top quark mass is, the larger
are the fluctuations, and hence the region where this
decrease takes place is shifted to larger Higgs masses. For a Higgs mass
beyond this region the survival of the BAU is ruled out,
irrespectively of its
initial value $B_{T_c}$ immediately after the phase transition. Let us
assume that this initial value is such that in order
to explain the present day BAU
we have to demand $B_0/B_{T_c}\simgt 10^{-5}$ \cite{Boch} (we
see, however, from Fig.~5 that this value is not important since a change by
many orders of magnitude alters the upper bound by only a few GeV).
For $m_t=150$ GeV we obtain $m_H\simlt 60$ GeV, for $m_t=174$ GeV
the upper bound is at 65 GeV, and for $m_t=200$ GeV the BAU survives
if $m_H\simlt 71$ GeV. At any rate, for all physical choices of
the parameters $m_t$ and $B_0/B_{T_c}$ the upper bound for the
Higgs mass is found in the range between 60 to 75 GeV.

The calculation without fermions leads to a qualitatively similar picture,
but the erasure of the asymmetry happens
already at much lower Higgs masses. Assuming $B_0/B_{T_c}\ge 10^{-5}$
we would obtain an upper bound for $m_H$ of only
49 GeV being close to the bound found previously by
Bochkarev and Shaposhnikov \cite{Shap,Boch}.
The large difference between this value and
the bound of 65 GeV which we obtain with fermion fluctuations for
$m_t=174$ GeV again confirms their significance.

\section{Summary and Conclusions}
\setcounter{equation}{0}

The present paper investigates the fate of the baryon number asymmetry
in the Universe (BAU) after the electroweak phase transition ($T<T_c$).
It is assumed that the asymmetry as such originates from baryogenic
processes before or during the phase transition with a net result
of $B+L\ne 0$ and $B-L=0$. It is furthermore assumed that in the broken
phase ($T<T_c$) the Minimal Standard Model with one Higgs doublet
holds. Since the Standard Model does not conserve the baryon number
due to possible sphaleron transitions, today's existence of an
asymmetry of about $10^{-10}$ baryons per photon implies certain
dynamic conditions right after the phase transition which prevent
too fast a ``wash-out'' of the baryon number \cite{Kuzmin}. In the
present paper the baryon number transition rate is evaluated in the one-loop
approximation around the classical sphaleron solution (hedgehog). The higher
loop effects are partially taken care of by an exact treatment of the
``Debye mass'' terms, $\sim \Phi^2T^2$.  It is assumed that
the Langer--Affleck formula holds and that no baryons are generated in
the broken phase.

For all temperatures below the critical temperature $T_c$ the
one-loop calculations are performed numerically in the limit
of vanishing Weinberg angle. In fact, the baryon number transition
rate depends on the classical sphaleron energy, the determinant of
the fermionic fluctuations, the determinant of the non-zero bosonic
fluctuations, the energy of the negative mode and the normalization factors
of the zero modes. While the sphaleron energy and the zero and negative
bosonic modes have been calculated previously in the literature
\cite{Arnold}, the evaluation of non-zero bosonic modes has been
performed only in the high temperature limit with somewhat controversial
results \cite{McLetal,Baacke}. In this context the present paper shows
the first calculation of the boson determinants for finite
temperatures  (the fermion determinant at arbitrary temperatures was
previously computed by the same authors \cite{rub}).
It turns out that all above contributions to the transition rate are
more or less equally important and must be evaluated at finite
temperatures in order to obtain, within the given conceptual frame, an
accurate calculation \cite{rub2}.

The actual numbers basically depend only on one unknown parameter, namely
the mass of the Higgs boson $m_H$. In fact, the dependence of the
baryon number transition rate on the Higgs mass is extremely strong.
Both bosonic and fermionic fluctuations
above the sphaleron barrier help to preserve the baryon asymmetry
in the Universe. They prevent a fast erasure of
the baryon excess provided the mass of the Higgs boson is less than some
upper bound,
while for larger Higgs masses the sphaleron transition rate becomes
large and the asymmetry would be eliminated. The value of this upper bound
depends on the mass of the top quark, ranging from about 60 GeV for
$m_t=150$ GeV to 71 GeV for $m_t=200$ GeV. These results are obtained
in the Minimal Standard Model with only one Higgs
doublet. They assume a theoretical frame characterized by
the applicability of the Langer--Affleck formula and the restriction to
one-loop calculations, with a partial resummation of higher orders.
These assumptions are only justified if the baryon number violating
processes do not happen immediately after the electroweak phase transition
where the loop expansion breaks down. This means the position of the
maximum of the transition rate $\gamma$ should be not too close
to the critical temperature. Moreover the quantum
loop contributions at the maximum should be small compared to the classical
terms. We find that both conditions are well fulfilled for Higgs masses
$m_H\simgt$ 100 GeV while this is not the case for small Higgs masses
below about 60 GeV. Those Higgs masses are, however, ruled out by experiment
\cite{Wyatt,Sopczak}.
Thus we arrive at the following conclusion: If the Higgs mass is in the range
between about 60 and 100 GeV, the Minimal Standard Model could be able to
account for the survival of the BAU, either within the formal framework we
used and a suitable top quark mass or by effects outside
this formalism, e.g.~higher loop contributions. If it is found above 100 GeV,
there is only a little chance to explain the present BAU within the MSM
since in this case the application of our framework is rather safe and predicts
the complete erasure of the BAU. A possible escape could be an extended model
with two Higgs doublets, following from supersymmetric models.

\medskip
{\bf Acknowledgements:}
We are grateful to A.Bochkarev, W.Buchm\"uller, Z.Fodor, L.McLerran,
and M.Shaposhnikov for very useful discussions and comments. We thank
P.Pobylitsa for his help on various stages of this work.
The work has been sponsored in part by the Deutsche
Forschungsgemeinschaft and by the International Science Foundation under
Grant No.~R2A000. D.D.~acknowledges the support of the Alexander von
Humboldt Foundation.

\newpage
\appendix
\section{A}
\setcounter{equation}{0}

In this appendix we describe how we solve numerically the eigenvalue
problems
\beq
{\cal K}_{\rm bos} \Psi_{\rm bos}=\omega^2\Psi_{\rm bos}\, ,
\hspace{2cm}
{\cal K}_{\rm FP} \Psi_{\rm FP}=\omega^2\Psi_{\rm FP}
\hspace{1cm}\la{eigenv}\eeq
for the boson fluctuation and the Faddeev--Popov operators.
We construct a finite basis for the fluctuations in which the operators
can numerically be diagonalized. Partially this technique has been
developed in the context of the chiral quark model
\cite{Kahana,Meissner,Sieber} and employed for the
diagonalization of the fermionic fluctuation matrix \cite{rub}, which
ensures consistency between the calculations of bosonic and fermionic
loop corrections.

The fluctuation vector $\Psi_{\rm bos}$ consists of nine
gauge field components $a_i^a$ ($a,i=1,\ldots,3$) and four
Higgs field components $\vp_\mu$ ($\mu =1,\ldots,4$) while
$\Psi_{\rm FP}$ contains only three components which we
denote by $a_0^a$. Hence the eigenvalue equations read:
\beq
\matr{{\cal G}_{ij}^{ab}}{{\cal W}_{i\nu}^a}{{\cal W}_{j\mu}^b}
{{\cal H}_{\mu\nu}}\vect{a_j^b}{\vp_\nu} = \omega^2
\vect{a_i^a}{\vp_\mu}\, , \hspace{1cm}
{\cal F}^{ab} a_0^b=\omega^2 a_0^a\, , \hspace{1cm}
\la{eigmat}\eeq
where the matrix elements of the fluctuation matrix ${\cal K}_{\rm bos}$
are given by (see \eq{Kfluc})
\begin{eqnarray}
{\cal G}_{ij}^{ab} &=& \delta_{ij}\delta^{ab}\left(-\partial^2
   + A_k^c A_k^c +\fracsm{1}{4}\Phi_\mu\Phi_\mu\right)
   + 2\ee^{abc}F_{ij}^c  \nonumber \\
   && \quad {}+\delta_{ij}\left(\ee^{abc}(\partial_k A_k^c)
    +2\ee^{abc}A_k^c\partial_k - A_k^a A_k^b\right)\, , \nonumber \\
{\cal H}_{\mu\nu} &=& \delta_{\mu\nu} \left(-\partial^2
  +\fracsm{1}{4} A_i^a A_i^a + \fracsm{1}{4}\Phi_\rho\Phi_\rho
  + \fracsm{1}{8}\nuH^2(\Phi_\rho\Phi_\rho-4)\right)\nonumber \\
  &&\quad {}+\eta^a_{\mu\nu} A^a_i\partial_i +\fracsm{1}{2} \eta^a_{\mu\nu}
   (\partial_i A^a_i) +\fracsm{1}{4} (\nuH^2-1)\Phi_\mu\Phi_\nu \, ,
  \nonumber \\
  {\cal W}_{i\nu}^a&=&
   \fracsm{1}{2}\Phi_\nu A^a_i -\fracsm{1}{2}\ee^{abc}
  \eta^c_{\mu\nu}\Phi_\mu A^b_i -\eta^a_{\mu\nu}(\partial_i
  \Phi_\mu)\, , \la{matrelements}
\end{eqnarray}
and the Faddeev--Popov operator by (see \eq{kFP})
\beq
{\cal F}^{ab}=\delta^{ab}\left(-\partial^2+ A_i^c A_i^c
+\fracsm{1}{4}\Phi_\mu\Phi_\mu\right)
+ \ee^{abc}(\partial_i A_i^c) +2\ee^{abc}A_i^c\partial_i-A_i^a A_i^b\, .
\la{fadpovmat}\eeq
For the static classical sphaleron fields we assume the spherically
symmetric hedgehog ansatz:
\begin{eqnarray}
A_i^a({\bf r}) &=& \ee_{aij}n_j\frac{1-A(r)}{r}
+ (\delta_{ai}-n_an_i)\frac{B(r)}{r}
+ n_an_i\frac{C(r)}{r}\,,
\nonumber \\
\Phi_i({\bf r}) &=& 2\,n_i\,G(r)\, ,\quad\quad
\Phi_4({\bf r}) = 2\,H(r)\, .
\la{HH}\end{eqnarray}
with the given five radial functions $A(r)$, $B(r)$, $C(r)$, $H(r)$,
$G(r)$. In principle one of these functions can be eliminated by a
gauge transformation. Although this would lead to a significant
simplification of the numerics we will not perform this step but rather
stick to the general gauge with five functions since our
numerical procedure only works if the classical fields are continuous
functions at zero and infinity, i.e.~that they take their vacuum values
there. This is only possible for a non-vanishing $C$ field, which
increases the numerical effort, but on the other hand allows
to check the invariance of all quantities under gauge transformations.

To exploit the spherical symmetry and to construct a finite basis in
which
${\cal K}_{\rm bos}$ and ${\cal K}_{\rm FP}$ can numerically be
diagonalized we consider a four ($=3+1$) dimensional reducible
$SU(2)$ representation with generators:
\beq
S_1=\left(\begin{array}{cccc}
0& 0& 0& 0\\ 0& 0& -i& 0\\ 0& i& 0& 0\\ 0& 0& 0& 0 \end{array}\right),\;
S_2=\left(\begin{array}{cccc}
0& 0& i& 0\\ 0& 0& 0& 0\\ -i& 0& 0& 0\\ 0& 0& 0& 0 \end{array}\right),\;
S_3=\left(\begin{array}{cccc}
0& -i& 0& 0\\ i& 0& 0& 0\\ 0& 0& 0& 0\\ 0& 0& 0& 0 \end{array}\right).
\la{SUgen}\eeq
 A basis of (``spin''-) eigenstates $\ket{S\,S_3}$ of ${\bf S}^2
=S_1^2+S_2^2+S_3^2$ and $S_3$ is given by
\beq
\ket{0\,0}=\vectf{0}{0}{0}{1},\;\ket{1\,1}=\frac{1}{\sqrt{2}}
\vectf{-i}{1}{0}{0}
,\; \ket{1\,0}=\vectf{0}{0}{i}{0},\; \ket{1\,-\!1}=\frac{1}{\sqrt{2}}
\vectf{i}{1}{0}{0}.\la{vectors}\eeq
The indices $i,j$ will always refer to coordinates of these
spin eigenstates, for example $\ket{1\,1}_{i=1}=-i/\sqrt{2}$,
$\ket{0\,0}_{i=4}=1$.
Similarly we define a four dimensional $SU(2)$ ``isospin''
representation; the operators $T_1$, $T_2$ and $T_3$
and the eigenstates $\ket{T\,T_3}$ look exactly as the corresponding
ones of the spin representation. Here the coordinates are referred
to by the indices $a,b$ and $\mu,\nu$. Moreover we use the basis
$\ket{L\,L_3}$ of the angular momentum operator to describe the spherical
space dependence of the fluctuations, with the property
\beq
\langle\Omega|L\,L_3\rangle=i^L\,Y_{L\,L_3}(\Omega)\, .
\la{harm}\eeq
The ``grand-spin''
operator defined by ${\bf K}={\bf J}+{\bf T}={\bf L}+{\bf S}+{\bf T}$
commutes with the fluctuation operators of \eq{eigenv}.
Therefore eigenstates of ${\bf K}^2$ and $K_3$ form a proper basis
for the diagonalization procedure.
We couple the eigenstates $\ket{L\,L_3}$, $\ket{S\,S_3}$ and
$\ket{T\,T_3}$ to eigenstates of ${\bf K}^2$ and $K_3$:
\beq
\ket{K,\,K_3;\,T,\,J,\,S,\,L}_i^a
=\sum_{L_3,S_3,J_3,T_3}
C^{K\,K_3}_{J\,J_3,\,T\,T_3} \, C^{J\,J_3}_{L\,L_3,\,S\,S_3}\;
 \ket{S\,S_3}_i \, \ket{T\,T_3}^a \, \ket{L\,L_3}\, .
\la{clebsh} \eeq
For the vacuum fluctuation operator, i.e.~in the case of no external field,
we can solve the eigenvalue problem
analytically; the dependence of the fluctuations on the radial
coordinate $r$ is in this case given by spherical Bessel functions.
We take
these solutions as the basis for a numerical diagonalization of $\cal K$
in the non-vacuum case. To this end we define states
$\ket{\,p;\,K,\,K_3;\,T,\,J,\,S,\,L}$ by
\beq
\langle{\bf r}\,\ket{\,p;\,K,\,K_3;\,T,\,J,\,S,\,L}={\cal N}\,j_L(pr)\,
    \langle\Omega\ket{K,\,K_3;\,T,\,J,\,S,\,L}\,.
\la{radbas}\eeq
Here the momentum $p$ is a continuous variable,
and $\cal N$ is a normalization factor specified below.
In order to get a finite basis we have to discretize the momentum and
to restrict its allowed values
to a finite number. With large enough numerical box parameters $R$ and
$P_{\rm max}$ we demand:
\beq
j_I(p_n^I R)=0\, , \hspace{3cm} p_n^I\le P_{\rm max}
\la{discr}\eeq
where
\beq
I=I(K,J,S)=\cases{K & for $S=0$ \cr J & for $S=1$ \cr}\ .
\eeq
In the case $S=T=1$ we have three discretization conditions for fixed
grand-spin $K$ instead of one, yielding
three sets of momenta $p_n^{K+1}$, $p_n^K$, and $p_n^{K-1}$.
This extension of the usual construction \cite{Meissner}
is necessary to ensure the orthogonality of the basis states;
it has already been used and checked in \cite{Sieber}.
We obtain
\begin{eqnarray}
&&{\cal N}_n^I{\cal N}_m^I\int_0^R dr\,r^2\;j_I(p_n^Ir)\,j_I(p_m^Ir)
\nonumber \\
&&\hspace{2cm}={\cal N}_n^I{\cal N}_m^I
\int_0^R dr\,r^2\;j_{I\pm 1}(p_n^Ir)\,j_{I\pm 1}(p_m^Ir)=\delta_{nm}\, ,
\la{ortho}\end{eqnarray}
if the normalization factor is chosen as
\beq
{\cal N}_n^I=\sqrt{\frac{2}{R^3}}\,\left|j_{I\pm 1}(p_n^I R)
\right|^{-1}\,;
\la{normfac}\eeq
hence our states are orthonormal:
\begin{eqnarray}
&&\langle p_n^I;\,K,\,K_3;\,T,\,J,\,S,\,L\,
  \ket{p_m^{I'};\,K',\,K_3';\,T',\,J',\,S',\,L'}\hspace{3cm}
\nonumber \\
&&\hspace{5cm}=
\delta_{nm}\delta_{KK'}\delta_{K_3K_3'}\delta_{TT'}\delta_{JJ'}
  \delta_{SS'}\delta_{LL'}\, .
\la{orthonorm}\end{eqnarray}
For fixed values of $K=0,1,2,\ldots$ and $K_3=-K,\ldots,+K$ we can
write down the following set of basis states for the fluctuations:
\begin{eqnarray}
\Psi_{\rm bos}^{1,\,\alpha}({\bf r})\!\!
&=&\hspace{-0.3cm}\vectt{a_i^a({\bf r})}{\vp_a({\bf r})}{\vp_4({\bf r})}
=\vectt{\langle{\bf r}\,\ket{\,p_n^J;\,K,\,K_3;\,1,\,J,\,1,\,L}_i^a}
{0}{0}
\rdef\vectt{\langle{\bf r}\,
\ket{\widetilde{\Psi}_{\rm bos}^{1,\,\alpha}}_i^a}
{0}{0}\nonumber \\
&&{\rm for}\; J=K-1,K,K+1;\; L=J-1,J,J+1;\; n=1,\ldots,N(J); \nonumber \\
\noalign{\vspace{0.4cm}}
\Psi_{\rm bos}^{2,\,\alpha}({\bf r})\!\!
&=&\hspace{-0.3cm}\vectt{a_i^a({\bf r})}{\vp_a({\bf r})}{\vp_4({\bf r})}
=\vectt{0}{\langle{\bf r}\,\ket{\,p_n^K;\,K,\,K_3;\,1,\,L,\,0,\,L}_4^a}
{0}\rdef
\vectt{0}{\langle{\bf r}\,\ket{\widetilde{\Psi}_{\rm bos}^{2,\,\alpha}}_4^a}
{0}\nonumber \\
&&{\rm for}\quad L=K-1,K,K+1;\quad n=1,\ldots,N(K); \nonumber \\
\noalign{\vspace{0.4cm}}
\Psi_{\rm bos}^{3,\,\alpha}({\bf r})\!\!
&=&\hspace{-0.3cm}\vectt{a_i^a({\bf r})}{\vp_a({\bf r})}{\vp_4({\bf r})}
=\vectt{0}{0}{\langle{\bf r}\,
\ket{\,p_n^K;\,K,\,K_3;\,0,\,K,\,0,\,K}_4^4}
\rdef
\vectt{0}{0}{\langle{\bf r}\,\ket{\widetilde{\Psi}_{\rm bos}^{3,\,\alpha}}_4^4}
\nonumber \\
&&{\rm for}\quad n=1,\ldots,N(K); \nonumber \\
\noalign{\vspace{0.4cm}}
\Psi_{\rm FP}^\alpha({\bf r})\!\!&=&\hspace{-0.2cm} a_0^a({\bf r})
=\langle{\bf r}\,\ket{\,p_n^K;\,K,\,K_3;\,1,\,L,\,0,\,L}_4^a
\rdef\langle{\bf r}\,\ket{\widetilde{\Psi}_{\rm FP}^\alpha}_4^a\nonumber \\
&&{\rm for}\quad L=K-1,K,K+1;\quad n=1,\ldots,N(K); \la{basis}
\end{eqnarray}
where $N(I)$ is the number of allowed momenta $p_n^I$, see \eq{discr}.
The index $\alpha$ enumerates the basis states of the three groups
for the fluctuations and of the Faddeev--Popov matrix.
For $K=0,1$ not all of these basis states exist. The
total number of states for $\Psi_{\rm bos}$ is given by
$3N(K+1)+7N(K)+3N(K-1)$ for fixed $K>1$ and fixed $K_3$,
for $K=1$ it is
$3N(2)+7N(1)+N(0)$ and $3N(1)+2N(0)$ for $K=0$. For the
Faddeev--Popov matrix we have $3N(K)$ basis vectors for $K>0$ and $N(0)$
for $K=0$.

We show below that due to the spherical symmetry of the
sphaleron the operator ${\cal K}_{\rm bos}$
is block diagonal in $K$ and $K_3$, i.e.~basis states with
different $K$ or $K_3$ do not mix. Moreover the blocks for different
$K_3$ and the same $K$ are identical, so that for each $K$ only one
matrix has to be diagonalized, and the resulting
eigenvalues are $(2K{+}1)$--fold degenerate. The dimension of this matrix
is given by the number of the above basis states. The same holds
for the Faddeev--Popov operator ${\cal K}_{\rm FP}$.

The remaining task is to calculate the matrix elements of the
operators in the basis \ur{basis};
i.e.~if $|\,\Psi_{\rm bos}^{\lambda_1,\,\alpha_1}\rangle$ and
$|\,\Psi_{\rm bos}^{\lambda_2,\,\alpha_2}\rangle$
are basis states given by \eq{basis}, we need to know
the element
$\langle\Psi_{\rm bos}^{\lambda_1,\,\alpha_1}|{\cal K}_{\rm bos}|
\Psi_{\rm bos}^{\lambda_2,\,\alpha_2}\rangle$.
For this purpose we have to express the
matrices in terms of spherical tensor operators so that the spherical
part of the matrix elements can be evaluated analytically. Apart from
$\bf S$ and $\bf T$ given in \eq{SUgen} we need
operators $\bf P^+$ and $\bf P^-$, acting in spin space, which we
define as
\beq
P^+_1=\left(\begin{array}{cccc}
0& 0& 0& 1\\ 0& 0& 0& 0\\ 0& 0& 0& 0\\ 1& 0& 0& 0 \end{array}\right),\;
P^+_2=\left(\begin{array}{cccc}
0& 0& 0& 0\\ 0& 0& 0& 1\\ 0& 0& 0& 0\\ 0& 1& 0& 0 \end{array}\right),\;
P^+_3=\left(\begin{array}{cccc}
0& 0& 0& 0\\ 0& 0& 0& 0\\ 0& 0& 0& 1\\ 0& 0& 1& 0 \end{array}\right),
\la{pplus}\eeq
and
\beq
P^-_1=\left(\begin{array}{cccc}
0& 0& 0& -i\\ 0& 0& 0& 0\\ 0& 0& 0& 0\\ i& 0& 0& 0 \end{array}\right),\;
P^-_2=\left(\begin{array}{cccc}
0& 0& 0& 0\\ 0& 0& 0& -i\\ 0& 0& 0& 0\\ 0& i& 0& 0 \end{array}\right),\;
P^-_3=\left(\begin{array}{cccc}
0& 0& 0& 0\\ 0& 0& 0& 0\\ 0& 0& 0& -i\\ 0& 0& i& 0 \end{array}\right).
\la{pminus}\eeq
It can easily be checked that like $\bf S$ these operators are
spherical vector operators, i.e.~$[K_i\, ,P^\pm_j]=i\,\ee_{ijk}
P^\pm_k$. Moreover we define two scalar operators by
\beq
I_S=\left(\begin{array}{cccc}
1& 0& 0& 0\\ 0& 1& 0& 0\\ 0& 0& 1& 0\\ 0& 0& 0& 0 \end{array}\right),
\quad\quad
i_S=\left(\begin{array}{cccc}
0& 0& 0& 0\\ 0& 0& 0& 0\\ 0& 0& 0& 0\\ 0& 0& 0& 1 \end{array}\right).
\la{ids}\eeq
In the isospin space we need the corresponding operators which we denote
by $\bf Q^\pm$ and $I_T$, $i_T$.

Using the relations
\begin{eqnarray}
\delta_{ij}&=& (I_S)_{ij}\; ,\quad \delta^{ab} = (I_T)^{ab}\;,\quad
\delta_{\mu\nu}=(I_T+i_T)_{\mu\nu}\,, \nonumber \\
\ee_{ijk}&=& i\,(S_k)_{ij} \; ,\quad
\ee^{abc}= i\,(T_c)^{ab}\; ,  \quad
\eta^a_{\mu\nu}=i\,(T_a+Q^-_a)_{\mu\nu}\,, \nonumber \\
A_k^aA_k^b&=&(A_k^cA_k^c I_T-A_k^cA_k^d T_dT_c)^{ab}\;, \quad
{\cal W}^a_{rb}=({\cal W}^k_{rk} I_T-{\cal W}^k_{rl} T_lT_k)^{ab}\,,
\nonumber \\
\Phi_\mu\Phi_\nu&=& (\Phi_k\Phi_k I_T + \Phi_4\Phi_4 i_T
 -\Phi_l\Phi_k T_lT_k+\Phi_k\Phi_4 Q^+_k)_{\mu\nu}\,, \nonumber \\
(P^+_r)_{i4}&=&(P^+_r)_{4i}=(iP^-_r)_{i4}=-(iP^-_r)_{4i}
=\delta_{ri}\, ,\nonumber \\
(Q^+_k)^{a4}&=&(Q^+_k)^{4a}=(iQ^-_k)^{a4}=-(iQ^-_k)^{4a}=\delta^{ka}
\la{relations}
\end{eqnarray}
we can rewrite the fluctuation matrices:
\begin{eqnarray}
{\cal G}^{ab}_{ij}&=&\biggl\{I_TI_S(-\partial^2+\fracsm{1}{4}
   \Phi_\mu\Phi_\mu)+I_S T_c T_d A_k^c A_k^d \nonumber \\
&&\hspace{0.5cm}{}+i\,T_c\left(2I_SA_k^c\partial_k
  + I_S(\partial_k A_k^c)+i\,\ee_{klm}F^c_{kl}S_m\right)
  \biggr\}^{ab}_{ij}\, ,\nonumber \\
{\cal H}_{\mu\nu} &=& \biggl\{i_S (I_T+i_T)\,
 \left(-\partial^2
  +\fracsm{1}{4} A_i^a A_i^a + \fracsm{1}{4}\Phi_\rho\Phi_\rho
  + \fracsm{1}{8}\nuH^2(\Phi_\rho\Phi_\rho-4)\right)\nonumber \\
  &&\hspace{0.5cm}{}+i_S(T_a+Q^-_a)\,i\,\left(A^a_i\partial_i+\fracsm{1}{2}
  (\partial_i A^a_i)\right)  \nonumber  \\
  &&\hspace{0.5cm}{}+i_S\,\fracsm{1}{4}(\nuH^2-1)\,
   \left(\Phi_k\Phi_k I_T + \Phi_4\Phi_4 i_T
     -\Phi_l\Phi_k T_lT_k+\Phi_k\Phi_4
        Q^+_k\right)\biggr\}_{44}^{\mu\nu}\, ,
    \nonumber \\
{\cal W}^a_{i\nu}&=&\biggl\{\fracsm{1}{2}\left({\cal W}^k_{rk}I_T
-{\cal W}^k_{rl}T_lT_k\right)\,(P^+_r+i P^-_r)\nonumber \\
&&\hspace{2cm}{}+\fracsm{1}{4}\,{\cal W}^k_{r4}\,
 (Q^+_k+i Q^-_k)\,(P^+_r+i P^-_r)\biggr\}^{a\nu}_{i4}\,, \nonumber \\
{\cal W}^b_{j\mu}&=&\biggl\{\fracsm{1}{2}\left({\cal W}^k_{rk}I_T
-{\cal W}^k_{rl}T_kT_l\right)\,(P^+_r-i P^-_r)\nonumber \\
&&\hspace{2cm}{}+\fracsm{1}{4}\,{\cal W}^k_{r4}\,
 (Q^+_k-i Q^-_k)\,(P^+_r-i P^-_r)\biggr\}^{\mu b}_{4j}\, , \nonumber \\
{\cal F}^{ab}&=& \biggl\{i_SI_T(-\partial^2+\fracsm{1}{4}\Phi_\mu\Phi_\mu)
    +i_ST_cT_d A_i^c A_i^d \nonumber \\
   &&\hspace{2cm}{}+i_ST_c\, i\,\bigl(2A^c_i\partial_i+(\partial_iA^c_i)\bigr)
  \biggr\}^{ab}_{44}\, . \la{matop}
\end{eqnarray}
Now we can find matrices ${\widetilde{\cal K}}_{\rm bos}$ and
${\widetilde{\cal K}}_{\rm FP}$ with the property
\begin{eqnarray}
\langle\Psi_{\rm bos}^{\lambda_1,\,\alpha_1}|\,{\cal K}_{\rm bos}\,|
\Psi_{\rm bos}^{\lambda_2,\,\alpha_2}\rangle &=&
\langle\widetilde{\Psi}_{\rm bos}^{\lambda_1,\,\alpha_1}|
\,{\widetilde{\cal K}}_{\rm bos}\,|
\widetilde{\Psi}_{\rm bos}^{\lambda_2,\,\alpha_2}\rangle \, ,\nonumber \\
\langle\Psi_{\rm FP}^{\alpha_1}|\,{\cal K}_{\rm FP}\,
|\Psi_{\rm FP}^{\alpha_2}\rangle
&=&\langle\widetilde{\Psi}_{\rm FP}^{\alpha_1}|\,
{\widetilde{\cal K}}_{\rm FP}\,
|\widetilde{\Psi}_{\rm FP}^{\alpha_2}\rangle   \la{newmat}
\end{eqnarray}
(for the definition of the states
$|\,\widetilde{\Psi}_{\rm bos}^{\lambda,\,\alpha}\rangle$ and
$|\,\widetilde{\Psi}_{\rm FP}^\alpha\rangle$ see \eq{basis}):
\begin{eqnarray}
{\widetilde{\cal K}}_{\rm bos}&=&I_TI_S(-\partial^2+\fracsm{1}{4}
   \Phi_\mu\Phi_\mu)+I_S T_c T_d A_k^c A_k^d
   +I_S T_c\Bigl(2A_k^c\,i\,\partial_k+i\,(\partial_k A_k^c)\Bigr)
   \nonumber \\
  &&{} -\ee_{klm}F^c_{kl}S_mT_c + i_S (I_T+i_T)\,
   \left(-\partial^2+\fracsm{1}{4} A_i^a A_i^a\right)\nonumber \\
  &&{} +i_SI_T\Bigl(\left(\fracsm{1}{4}
  +\fracsm{1}{8}\nuH^2\right)
   \Phi_4\Phi_4+\fracsm{3}{8}\nuH^2\Phi_k\Phi_k-\fracsm{1}{2}\nuH^2
   \Bigr)\nonumber \\
   &&{}+i_Si_T\Bigl(\left(\fracsm{1}{4}
   +\fracsm{1}{8}\nuH^2\right)
   \Phi_k\Phi_k+\fracsm{3}{8}\nuH^2\Phi_4\Phi_4-\fracsm{1}{2}\nuH^2
   \Bigr)\nonumber \\
   &&{}+i_S(T_a+Q^-_a)\left(A^a_i\,i\,\partial_i
   +\fracsm{1}{2}(i\,\partial_i A^a_i)\right)\nonumber\\
   &&{}+\fracsm{1}{4}(\nuH^2-1)\,
   \Bigl(i_SQ_k^+\Phi_k\Phi_4-i_ST_lT_k
   \Phi_k\Phi_l\Bigr)
   +I_TP_r^+{\cal W}^k_{rk} \nonumber \\
   &&{}-\fracsm{1}{2}\ee_{klm}{\cal W}^k_{rl}T_mP_r^-
   -\fracsm{1}{2}\left(
   {\cal W}^k_{rl}+{\cal W}^l_{rk}\right)T_lT_kP_r^+  \nonumber \\
   &&{}+\fracsm{1}{2}
   {\cal W}^k_{r4}\left(Q_k^+P_r^+-Q_k^-P_r^-\right)\, ,\nonumber \\
   {\widetilde{\cal K}}_{\rm FP}&=&
   i_SI_T(-\partial^2 +\fracsm{1}{4}\Phi_\mu\Phi_\mu)
    +i_ST_cT_d A_i^c A_i^d
   +i_S T_c\Bigl(2A_i^c\,i\,\partial_i+i\,(\partial_i A_k^c)\Bigr)\, .
   \nonumber \\
   \la{tildemat}
\end{eqnarray}
Our next step is to plug in the hedgehog ansatz \ur{HH} so that the
matrices
${\widetilde{\cal K}}_{\rm bos}$ and ${\widetilde{\cal K}}_{\rm FP}$ can be
expressed through the profile functions $A$, $B$, $C$, $H$, $G$, and
the spherical vector and scalar operators. With the hedgehog ansatz
and
\beq
\partial_i = n_i\,\fracsm{\partial}{\partial r}-\fracsm{i}{r}\,
   \ee_{ijk}n_jL_k
 \la{angular}\eeq
we obtain after a long and tedious calculation:
\begin{eqnarray}
{\widetilde{\cal K}}_{\rm bos}&=&\biggl(-\fracsm{\partial^2}{\partial r^2}
   -\fracsm{2}{r}\,\fracsm{\partial}{\partial r}+\fracsm{{\bf L}^2}{r^2}
   + G^2 + H^2+\fracsm{2}{r^2}\left((1-A)^2+B^2\right)\biggr)I_TI_S
   \nonumber \\
   &&{}+\fracsm{1}{r^2}\left(C^2-B^2-(1-A)^2\right)I_S\,
    ({\bf n}\cdot{\bf T})^2+\fracsm{2}{r^2}\,(1-A)\,I_S\,
    ({\bf T}\cdot{\bf L})
   \nonumber \\
   &&{}+\fracsm{i}{r^2}\left(2rC\fracsm{\partial}{\partial r}
    +rC'+C\right)I_S\,({\bf n}\cdot{\bf T}) +\fracsm{2B}{r^2}\,I_S\,
    \Bigl({\bf T}\cdot({\bf n}\times{\bf L})-i\,({\bf n}\cdot{\bf T})
    \Bigr)    \nonumber \\
   &&{}+\fracsm{2}{r^2}
   \Bigl(1-A^2-B^2+rA'+BC\Bigr)\,({\bf n}\cdot{\bf S})\,
      ({\bf n}\cdot{\bf T}) \nonumber \\
    &&{}+\fracsm{2}{r^2}\,(rB'-AC)\;{\bf n}\cdot({\bf S}
      \times{\bf T})-\fracsm{2}{r^2}\,(rA'+BC)\,({\bf S}\cdot{\bf T})
      \nonumber \\
   &&{}+\biggl(-\fracsm{\partial^2}{\partial r^2}
   -\fracsm{2}{r}\,\fracsm{\partial}{\partial r}+\fracsm{{\bf L}^2}{r^2}
   +\fracsm{1}{2r^2}\left((1-A)^2+B^2+\fracsm{C^2}{2}\right)\biggr)
    i_S(I_T+i_T) \nonumber \\
   &&{}+\biggl(H^2+\fracsm{1}{2}\nuH^2(H^2-1)
    +\fracsm{3}{2}\nuH^2G^2\biggr)i_SI_T\nonumber \\
   &&{}+\biggl(G^2+\fracsm{1}{2}\nuH^2(G^2-1)
    +\fracsm{3}{2}\nuH^2H^2\biggr)i_Si_T
    +(1-\nuH^2)\,G^2\,i_S\,({\bf n}\cdot{\bf T})^2
    \nonumber \\
   &&{}\fracsm{1}{r^2}\,(1-A)\,i_S\,({\bf T}+{\bf Q}^-)\cdot{\bf L}
    +\fracsm{i}{2r^2}\left(2rC\fracsm{\partial}{\partial r}
    +rC'+C\right)i_S\,{\bf n}\cdot({\bf T}+{\bf Q}^-) \nonumber \\
    &&{}+\fracsm{B}{r^2}\,i_S\,
    \Bigl(({\bf T}+{\bf Q}^-)\cdot({\bf n}\times{\bf L})
    -i\,{\bf n}\cdot({\bf T}+{\bf Q}^-)\Bigr)  \nonumber \\
   &&{}-(1-\nuH^2)\,HG\,i_S\,({\bf n}\cdot{\bf Q}^+)
    +\fracsm{1}{r}\,\left(GC+2rH'\right)\,I_T\,({\bf n}\cdot{\bf P}^+)
    \nonumber \\
   &&{}-\fracsm{1}{r}\,(G+GA-HB)\,({\bf T}\cdot{\bf P}^-)
   +\fracsm{1}{r}\,(H-HA-BG)\,{\bf n}\cdot({\bf T}\times{\bf P}^-)
   \nonumber \\
   &&{}+\fracsm{1}{r}\,(G+GA-HB+HC-2rG')\,({\bf n}\cdot{\bf T})\,
   ({\bf n}\cdot{\bf P}^-) \nonumber \\
   &&{}+\fracsm{1}{2r}\,(H-HA-BG)\,{\bf n}\cdot({\bf Q}^+\times{\bf P}^+
      -{\bf Q}^-\times{\bf P}^-) \nonumber \\
   &&{}-\fracsm{1}{2r}\,(G+GA-HB)\,({\bf Q}^+\cdot{\bf P}^+
     -{\bf Q}^-\cdot{\bf P}^-) \nonumber \\
   &&{}+\fracsm{1}{2r}\,(G+GA-HB+HC-2rG') \nonumber \\
   &&\hspace{3cm}\Bigl(({\bf n}\cdot{\bf Q}^+)
   \,({\bf n}\cdot{\bf P}^+)-({\bf n}\cdot{\bf Q}^-)\,
    ({\bf n}\cdot{\bf P}^-)
   \Bigr)\, ,\nonumber \\
{\widetilde{\cal K}}_{\rm FP}&=&\biggl(-\fracsm{\partial^2}{\partial r^2}
   -\fracsm{2}{r}\,\fracsm{\partial}{\partial r}+\fracsm{{\bf L}^2}{r^2}
   + G^2 + H^2+\fracsm{2}{r^2}\left((1-A)^2+B^2\right)\biggr)i_SI_T
   \nonumber \\
   &&{}+\fracsm{1}{r^2}\left(C^2-B^2-(1-A)^2\right)i_S\,
  ({\bf n}\cdot{\bf T})^2
    +\fracsm{2}{r^2}\,(1-A)\,i_S\,({\bf T}\cdot{\bf L})
   \nonumber \\
   &&{}+\fracsm{i}{r^2}\left(2rC\fracsm{\partial}{\partial r}
    +rC'+C\right)i_S\,({\bf n}\cdot{\bf T}) +\fracsm{2B}{r^2}\,i_S\,
    \Bigl({\bf T}\cdot({\bf n}\times{\bf L})
    -i\,({\bf n}\cdot{\bf T})\Bigr)\, ,
    \nonumber \\
    \la{hedgemat}
\end{eqnarray}
where we dropped the argument $r$ of the profile functions. Now it is
easy to see that all the spherical operators which turn up in
\eq{hedgemat} are scalar operators in the sense that they commute
with the grand-spin ${\bf K}^2$ and $K_3$ so that
\beq
\left[{\widetilde{\cal K}}_{\rm bos}\,,\,{\bf K}^2\right]=
\left[{\widetilde{\cal K}}_{\rm bos}\,,\,K_3\right]=0\,,\quad\quad
\left[{\widetilde{\cal K}}_{\rm FP}\,,\,{\bf K}^2\right]=
\left[{\widetilde{\cal K}}_{\rm FP}\,,\,K_3\right]=0\, .
\eeq
This is the reason why the matrices can be diagonalized in each $K$
sector separately, as mentioned above.

For the numerical diagonalization one has to evaluate these matrices
in the basis \ur{basis}. The radial part of a matrix element leads
to a numerical computation of a one dimensional
integral. The angular part, however, can be evaluated analytically.
The most direct and easiest way to do this is to employ the
Wigner--Eckart theorem and perform the summations over the
Clebsch--Gordan
coefficients with the help of a program like {\it Mathematica}.
Instead of writing down the complete result of this angular integration
we demonstrate the procedure with an example. We consider the operator
$\fracsm{1}{r}\,(GC+2rH')\,I_T\,({\bf n}\cdot{\bf P}^+)$
which is part of
${\widetilde{\cal K}}_{\rm bos}$ and calculate the matrix element
\[
\Bigl<p_n^{J};\,K,K_3;\,1,\,J,\,1,\,L\,\left|\,
  \fracsm{1}{r}\,(GC+2rH')\,I_T\,({\bf n}\cdot{\bf P}^+)\,\right|\,
 p_m^K;\,K,K_3;\,1,\,L',\,0,\,L'\Bigr>
\]\[
  ={\cal N}_n^{J}{\cal N}_m^K\int_0^R
 dr\;r^2\,j_L(p_n^{J}r)\,\fracsm{1}{r}\,(GC+2rH')\,j_{L'}(p_m^K)
   \hspace{2.0cm}
\]\beq
 \hspace{0.5cm}\cdot\,\Bigl<\,K,K_3;\,1,\,J,\,1,\,L\,\left|\,I_T\,
    ({\bf n}\cdot{\bf P}^+)\,\right|\,K,K_3;\,1,\,L',\,0,\,L'\,\Bigr>\, .
   \la{elesplit}
\eeq
The integral over the radial coordinate $r$ has to be evaluated numerically.
The angular matrix element is independent of $K_3$ so that we can put $K_3$
to zero. We obtain:
\begin{eqnarray}
&&\Bigl<\,K,0;\,1,\,J,\,1,\,L\,\left|\,I_T\,
({\bf n}\cdot{\bf P}^+)\,\right|\,K,0;\,1,\,L',\,0,\,L'\,\Bigr>
\nonumber\\
&&\hspace{1cm}=\sum_{L_3,S_3,J_3,T_3}\;\sum_{L'_3,T'_3}\;\sum_{m=-1}^1
\,(-1)^m\,
C^{K\,0}_{J\,J_3,\,1\,T_3}\,C^{J\,J_3}_{L\,L_3,\,1\,S_3}\,
C^{K\,0}_{L'\,L'_3,\,1\,T'_3}\quad\quad\quad \nonumber \\
&&\hspace{1.5cm}\cdot\,\langle\,1\,T_3\,|\,I_T\,|\,1\,T'_3\,\rangle\;
\langle\,1\,S_3\,|\,P^+_{(m)}\,|\,0\,0\,\rangle\;
\langle\,L\,L_3\,|\,n_{(-m)}\,|\,L'\,L'_3\,\rangle\, .\la{example}
\end{eqnarray}
Here the operators $P^+_{(m)}$ and $n_{(-m)}$ are spherical, not
cartesian components of ${\bf P}^+$ and ${\bf n}$. The matrix elements in
the last line can subsequently be evaluated using the Wigner--Eckart
theorem, e.g.
\beq
\langle\,1\,S_3\,|\,P^+_{(m)}\,|\,0\,0\,\rangle=\frac{1}{\sqrt{3}}\,
C^{1\,S_3}_{0\,0,\,1\,m}\,\langle\,1\,||\,{\bf P}^+\,||\,0\,\rangle\, .
\la{WigEck}\eeq
Hence for the calculation of the matrix element in \eq{elesplit} we need
to perform a numerical integration, a summation over Clebsch--Gordan
coefficients, and to know the following reduced matrix elements:
\begin{eqnarray}
\langle\,L\,||\,{\bf L}\,||\,L'\,\rangle&=&\delta_{LL'}
   \;\sqrt{L(L+1)(2L+1)}\,,\nonumber \\
\langle\,L\,||\,{\bf n}\,||\,L'\,\rangle&=&\delta_{|L-L'|,\,1}\;(-i)
   \;\sqrt{\fracsm{1}{2}(L+L'+1)}\,,\nonumber \\
\langle\,L\,||\,{\bf n}\times{\bf L}\,||\,L'\,\rangle&=&
  \cases{-(L-1)\sqrt{L} & for $L'=L-1$ \cr
          (L+2)\sqrt{L+1} & for $L'=L+1$ \cr
           0 & otherwise}\,,
  \nonumber\\
\langle\,S\,||\,{\bf S}\,||\,S'\,\rangle&=&\delta_{SS'}
   \;\sqrt{S(S+1)(2S+1)}\,,\nonumber \\
\langle\,S\,||\,I_S\,||\,S'\,\rangle&=&\sqrt{3}\,\delta_{S1}\,
    \delta_{S'1}\,, \nonumber \\
\langle\,S\,||\,i_S\,||\,S'\,\rangle&=&\delta_{S0}\,
    \delta_{S'0}\,, \nonumber \\
\langle\,S\,||\,{\bf P}^+\,||\,S'\,\rangle&=&-i\,\sqrt{3}\;
 (\delta_{S1}\,\delta_{S'0}+\delta_{S0}\,\delta_{S'1})\,,
 \nonumber \\
\langle\,S\,||\,{\bf P}^-\,||\,S'\,\rangle&=&-\sqrt{3}\;
   (\delta_{S1}\,\delta_{S'0}-\delta_{S0}\,\delta_{S'1})\,.\la{reduced}
\end{eqnarray}
For the correspoding reduced matrix elements of the isospin states one
simply has to replace $S$ by $T$ and $P$ by $Q$.

\section{B}
\setcounter{equation}{0}

We discuss here the spectral densities $\vrho(E)$ defined in \eq{dens},
especially their asymptotic behavior for large $E$.
It is easy to see that at large $E$ one can expand the spectral density
in a series
\beq
\vrho(E)\sim \vrho_\infty  +
\sum_{n=1}^\infty \vrho_{2n}E^{-2n}\;.
\la{rhoasym}\eeq
In order to calculate the values of the coefficients
\beq
\lim\limits_{E\to\infty }\vrho(E)=\vrho_\infty \qquad\mbox{and}\qquad
\lim\limits_{E\to\infty }\Bigl(\vrho(E)-\vrho_\infty \Bigr)E^2=\vrho_2
\la{rholim}\eeq
 we use the small $t$ expansion of
\begin{eqnarray}
F(t)&=&\Tr\Bigl(\exp[-t\K]-\exp[-t\K^{(0)}]\Bigr)=
\sum_{\SSS\rm discrete\atop levels}e^{-t\omega^2}
+\int_0^\infty  dE\,\vrho(E)\,e^{-tE^2}\nonumber\\
&=&at^{-1/2}+bt^{1/2}+ct^{3/2}+\cdots\;.
\la{Fexp}\end{eqnarray}
The corresponding coefficients can be easily calculated using gradient
expansion, following the renormalization procedure
\eqsss{Tr-small-t-asymptotics}{reno3} one can read off
\begin{eqnarray}
a_{\rm FP}&=&-\fracsm{3}{32\pi^{3/2}}\int d^3{\bf r}\,
(\bar\Phi^\dagger\bar\Phi-4)\;,\nonumber\\
a_{\rm bos}&=&-\fracsm{3}{32\pi^{3/2}}(4+\nuH^2)\int d^3{\bf r}\,
(\bar\Phi^\dagger\bar\Phi-4)\;,\nonumber\\
a_{\rm ferm}&=&-\fracsm{1}{8\pi^{3/2}}\,N_c\nut^2\int d^3{\bf r}\,
(\bar\Phi^\dagger\bar\Phi-4)
\la{aaa}\\
b_{\rm FP}&=&\fracsm{1}{16\pi^{3/2}}\int d^3{\bf r}\,
\Bigl(-\fracsm{1}{3}(\bar F^a_{ij})^2
+\fracsm{3}{16}(\bar\Phi^\dagger\bar\Phi-4)^2
+\fracsm{3}{2}(\bar\Phi^\dagger\bar\Phi-4)\Bigr)\;,\nonumber\\
b_{\rm bos}&=&\fracsm{1}{16\pi^{3/2}}\int d^3{\bf r}\,\Bigl(
\fracsm{41}{6}(\bar F^a_{ij})^2+\fracsm{3}{16}(4-3\nuH^2+\nuH^4)
(\bar\Phi^\dagger\bar\Phi-4)^2\;,\nonumber\\
&&\qquad\qquad\qquad{}+\fracsm{3}{4}(8-3\nuH^2+\nuH^4)
(\bar\Phi^\dagger\bar\Phi-4)\Bigr)\;,\nonumber\\
b_{\rm ferm}&=&\fracsm{1}{16\pi^{3/2}}\int d^3{\bf r}\,\Bigl(
\fracsm{1}{3}\,N_g(N_c+1)(\bar F^a_{ij})^2
+\fracsm{1}{8}\,N_c\nut^2(2\nut^2-\nuH^2)(\bar\Phi^\dagger\bar\Phi-4)^2
\;,\nonumber\\
&&\qquad\qquad\qquad{}+\fracsm{1}{2}\,N_c\nut^2(4\nut^2-\nuH^2)
(\bar\Phi^\dagger\bar\Phi-4)\Bigr)\;,
\la{bbb}\end{eqnarray}
where we made use of \eq{motion} to eliminate the term
$(\bar D_i\bar\Phi)^\dagger(\bar D_i\bar\Phi)$ in $b_{\rm bos}$
and $b_{\rm ferm}$.
In Section 7 we also need \cite{McLCar}
\begin{eqnarray}
c_{\rm bos}&=&
-\fracsm{1}{384\pi^{3/2}}\int d^3{\bf r}\,\biggl[2\nuH^2{\bar F}_{ij}^a
{\bar F}_{ij}^a+\fracsm{28}{15}\,\ee^{abc}{\bar F}_{ij}^a
{\bar F}_{jk}^b{\bar F}_{ki}^c \nonumber \\
&&\hspace{2.2cm}{}+\fracsm{1}{4}\,(-3\nuH^2+93)\,
(\bar\Phi^\dagger\bar\Phi)\,{\bar F}_{ij}^a{\bar F}_{ij}^a
+\fracsm{1}{8}\,(5\nuH^4-4\nuH^2+\fracsm{449}{5})\,
[\partial_i(\bar\Phi^\dagger\bar\Phi)]^2\nonumber \\
&&\hspace{2.2cm}{}+\fracsm{1}{2}\,(\nuH^4+28\nuH^2+\fracsm{31}{5})\,
(\bar\Phi^\dagger\bar\Phi)\,({\bar D}_i\bar\Phi)^\dagger
({\bar D}_i\bar\Phi)\nonumber \\
&&\hspace{2.2cm}{}+\fracsm{1}{32}\,(15\nuH^6+21\nuH^4+18\nuH^2+48)\,
(\bar\Phi^\dagger\bar\Phi-4)^3\nonumber \\
&&\hspace{2.2cm}{}+\fracsm{1}{8}\,(27\nuH^6+57\nuH^4+36\nuH^2+144)\,
(\bar\Phi^\dagger\bar\Phi-4)^2\nonumber \\
&&\hspace{2.2cm}{}+9\,(\nuH^6+2\nuH^4+\nuH^2+8)\,
(\bar\Phi^\dagger\bar\Phi-4)\biggr]\,.
\la{cbos}\end{eqnarray}
Using \ra{Fexp} one can show that
\beq
\lim\limits_{t\to 0}\sqrt{t}\,F(t)
=\fracsm{1}{2}\,\sqrt{\pi}
\,\vrho_\infty  \;,
\la{rhoa}\eeq
which implies
\beq
\vrho_\infty =2a/\sqrt{\pi}\;.
\la{rho0a}\eeq
To  calculate $\vrho_2$, let us introduce the function $R(E)$ such
that
\beq R'(E)=\vrho(E)\quad\mbox{and}\quad R(0)=n_{\SSS
D}=\mbox{number of discrete levels}\;, \la{RnD}\eeq $n_{\SSS D}^{\rm
FP}=0\,,\ n_{\SSS D}^{\rm bos}=7\,, \ n_{\SSS D}^{\rm ferm}=1\,$.
Using partial integration, we can write
\begin{eqnarray}
\lefteqn{
\frac{\sqrt{t}\,F(t)-a}{t}=\frac{1}{\sqrt{t}}\,\biggl(
\sum_{\SSS\rm discrete\atop levels}e^{-t\omega^2}+\int_0^\infty  dE\,
\Bigl(\vrho(E)-\vrho_\infty \Bigr)\,e^{-tE^2}\biggr)}\nonumber\\
&&\hspace{-5mm}
=\frac{1}{\sqrt{t}}\,\biggl(\sum_{\SSS\rm discrete\atop levels}e^{-t\omega^2}
+\Bigl(R(E)-\vrho_\infty E\Bigr)\,e^{-tE^2}\biggr|^\infty _0\biggr)
+2\sqrt{t}\int_0^\infty  dE\,
\Bigl(R(E)-\vrho_\infty E\Bigr)\,E\,e^{-tE^2}\nonumber\\
&&\hspace{-5mm}
=\frac{1}{\sqrt{t}}\sum_{\SSS\rm discrete\atop levels}(e^{-t\omega^2}-1)
+2\sqrt{t}\int_0^\infty  dE\,
\Bigl(R(E)-\vrho_\infty E\Bigr)\,E\,e^{-tE^2}\;.
\la{rhoR}\end{eqnarray}
The first term vanishes with $t\to0$, and for the second one we can use
the same way as above with $\Bigl(R(E)-\vrho_\infty E\Bigr)\,E$ instead of
$\vrho(E)$ and find that
\beq
\lim\limits_{E\to\infty }
\Bigl(R(E)-\vrho_\infty E\Bigr)\,E=\frac{b}{\sqrt{\pi}}\;.
\la{limRb}\eeq
Using l'Hospital's rule we finally find
\beq
\vrho_2=\lim\limits_{E\to\infty }\Bigl(\vrho(E)-\vrho_\infty \Bigr)\,E^2
=-\frac{b}{\sqrt{\pi}}\;.
\la{rho2b}\eeq
{}From \eq{limRb} we can deduce another interesting result;
$\lim\limits_{E\to\infty }\Bigl(R(E)-\vrho_\infty E\Bigr)=0$ yields
\beq
\int_0^\infty  dE\,\Bigl(\vrho(E)-\vrho_\infty \Bigr)=\lim\limits_{E\to\infty}
\Bigl(R(E)-\vrho_\infty E-R(0)\Bigr)=-R(0)=-n_{\SSS D}\;.
\la{intrnd}\eeq

Finally we show that the contributions $E^{\rm small}_{\cdots}$ to the
transition rate $\gamma$ are finite in the high temperature limit,
i.e.~$q\to0$.
In the fermionic case we obtain by partial integration
\begin{eqnarray}
E_{\rm ferm}^{\rm small}\Bigr|_{qm_W}\!\!\!&=&\!\!\!
-\fracsm{1}{\beta}\int_0^\infty dE\,
\Bigl(\vrho^{\rm ferm}(E)-\vrho_\infty^{\rm ferm}\Bigr)\,
\ln(1+e^{-\beta qm_WE})-\fracsm{1}{\beta}\,n_{\SSS D}^{\rm ferm}\ln 2
\quad\nonumber\\
&=&\!\!\!\fracsm{1}{\beta}\,\Bigl(
\underbrace{R^{\rm ferm}(0)-n_{\SSS D}^{\rm ferm}}_{0}\Bigr)
\,\ln 2-qm_W\int_0^\infty dE\,\frac{R^{\rm ferm}(E)-\vrho_\infty^{\rm ferm}E}
{1+e^{\beta qm_WE}} \nonumber \\
&\stackrel{q\to0}{\longrightarrow}&0\;.\la{fermhT}
\end{eqnarray}
For $E_{\rm bos}^{\rm small}$ and $E_{\rm FP}^{\rm small}$ we obtain
\begin{eqnarray}
E_{\cdots}^{\rm small}\Bigr|_{qm_W}
&=&\!\!\pm\fracsm{1}{\beta}\int_0^\infty dE\,
\Bigl(\vrho^{\cdots}(E)-\vrho^{\cdots}_\infty\Bigr)
\,\ln\frac{1-e^{-\beta qm_WE}}{\beta qm_W}\nonumber\\
&\stackrel{q\to0}{\longrightarrow}&
\pm\fracsm{1}{\beta}\int_0^\infty dE\,
\Bigl(\vrho^{\cdots}(E)-\vrho^{\cdots}_\infty\Bigr)\,\ln E
\qquad\mbox{(finite)}
\la{boshT}\end{eqnarray}
The last line holds
since this is true for the integral $\int_0^E$ with arbitrary
upper bound $E$, and the rest $\int_E^\infty$ vanishes with growing $E$
due to the behavior of $\vrho(E)-\vrho_\infty$, \eq{rho2b}.

\newpage
\vspace*{\fill}
$$
\begin{array}{|c|c|c|c|c|c|}
\hline
R & 10  & 10 & 10 & 12 & 14 \\
\hline
P_{\rm max} & 12 & 14 & 16 & 16 & 16 \\
\hline
\hline
E_{\rm bos}^{\rm conv}(\Lambda=4)/m_W &-6.25 & -6.29 & -6.29 & -6.28 &
-6.28 \\
\hline
\end{array}
$$
\par\bigskip\noindent
{\bf Tab.~1:}
$E_{\rm bos}^{\rm conv}(\Lambda=4)$ for different values of
the numerical parameters $R$ and $P_{\rm max}$. The result shows that
$R=12$ and $P_{\rm max}=16=4\,\Lambda$ are large enough
to ensure that the continuum limit is reached. The mass parameters are
$m_H=m_W=$ 83 GeV, $m_t=$ 174 GeV, $\nu_{\rm ren}=2.02$.
\vspace*{\fill}
$$
\begin{array}{|c|c|c|c|c|c|c|c|}
\hline
\Lambda & 2 & 3 & 4  & 4.5  & 5   & 5.5 & 6  \\
\hline
E_{\rm bos}^{\rm conv}(\Lambda)/m_W & -6.85 & -6.47 & -6.28 & -6.22
&-6.18 & -6.14 & -6.11 \\
\hline
\end{array}
$$
\par\bigskip\noindent
{\bf Tab.~2:}
$E_{\rm bos}^{\rm conv}(\Lambda)$ for various values of $\Lambda$.
Using the behaviour
$E_{\rm bos}^{\rm conv}(\Lambda)=E_{\rm bos}^{\rm ren} + b/\Lambda^2$
for large $\Lambda$, we find by extrapolation
the continuum limit $E_{\rm bos}^{\rm ren}=-5.95\,m_W$. The mass parameters
are $m_H=m_W=$ 83 GeV, $m_t=$ 174 GeV, $\nu_{\rm ren}=2.02$.
\vspace*{\fill}
$$
\begin{array}{|c||c|c|c|c|c|c|c|c|}
\hline
m_H\,{\rm [GeV]} & 50  & 66  & 66  & 66  & 83  & 100 & 150 & 350 \\
\hline
m_t\,{\rm [GeV]} & 174 & 150 & 174 & 200 & 174 & 174 & 174 & 174 \\
\hline
\hline
\nu_{\rm ren} & 2.07 & 1.74 & 2.05 & 2.36 & 2.02 & 1.98 & 1.76 & 2.11 \\
\hline
\hline
E_{\rm bos}^{\rm ren}/m_W  & -9.74 & -5.40 & -7.22 & -9.32
& -5.95 & -5.09 & -3.64 & -5.69 \\
\hline
E_{\rm FP}^{\rm ren}/m_W  &  1.36 &  0.48 &  0.75 &  1.10
& 0.46 &   0.29 & 0.08 & -0.01 \\
\hline
(E_{\rm bos}^{\rm ren}+E_{\rm FP}^{\rm ren})/m_W
 & -8.38 & -4.92 & -6.47 & -8.22 & -5.49 & -4.80 & -3.56 & -5.69 \\
\hline
\hline
E_{\rm ferm}^{\rm ren}/m_W  & 26.94 & 10.35 & 16.91 &
26.81  & 12.11&  9.40 & 5.87 & 4.05 \\
\hline
E_{\rm class}/m_W  & 96.94 & 99.60 & 99.60 &  99.60 & 101.94 & 104.08
& 109.27 & 121.67 \\
\hline
\end{array}
$$
\par\bigskip\noindent
{\bf Tab.~3:}
The renormalized non-thermal energy of the boson fluctuations
$E_{\rm bos}^{\rm ren}$, of the Faddeev--Popov operator
$E_{\rm FP}^{\rm ren}$, and the sum of both for various values of
the Higgs and the top quark mass. For comparison the classical
sphaleron energy $E_{\rm class}$ and the fermionic non-thermal energy
$E_{\rm ferm}^{\rm ren}$ are included. The renormalization scale is
determined according to \eq{nurenfix}.
\vspace*{\fill}
\newpage
\vspace*{\fill}
$$
\begin{array}{|c|c|c|c|c|c|}
\hline
E_a & 2.0 & 3.0 & 4.0  & 6.0  & 8.0    \\
\hline
E_b & 1.0 & 1.5 & 2.0  & 3.0  & 4.0    \\
\hline
\hline
{\rm 1st\;line\;of\;\eq{numkapp}\;(sum)} &  2.54 & 10.95 & 21.93 &
49.95 & 83.95 \\
\hline
{\rm 2nd\;and\;3rd\;lines\;of\;\eq{numkapp}\;(integrals)} &  3.64
& -4.20 & -15.07 & -43.10 & -77.13 \\
\hline
\hline
\beta_c E^{\rm small}_{\rm bos}(T_c) & 6.18 & 6.74 & 6.85 &
6.85 & 6.82 \\
\hline
\end{array}
$$
\par\bigskip\noindent
{\bf Tab.~4:}
$\beta_c E^{\rm small}_{\rm bos}(T_c)$
and its contributions for several
values of the numerical parameters $E_a$ and $E_b$. The contributions
strongly depend on $E_a$ and $E_b$, but $E^{\rm small}_{\rm bos}$
is very stable in the range $3\le E_a\le 8$.
\vspace*{\fill}
$$
\begin{array}{|c||c|c|c|}
\hline
m_H\,{\rm [GeV]} & 66 & 83 & 125  \\
\hline
\hline
\ln\chi_{\rm bos} & -11.66 & -6.85 & -1.96 \\
\hline
\ln\tilde\chi_{\rm bos} & -12.91 & -7.66 & -2.27 \\
\hline
\end{array}
$$
\par\bigskip\noindent
{\bf Tab.~5:}
Exact and approximate results for the high temperature limit of the boson
fluctuation determinant for various $m_H$. The exact values
$\ln\chi_{\rm bos}$ are determined by a summation over the spectrum of
eigenvalues of the fluctuation operator, the approximate values
$\ln\tilde\chi_{\rm bos}$ are
obtained with the DPY-method \cite{DPY}.
One finds an accuracy of about 10 to 15$\%$.
\vspace*{\fill}
\end{document}